\documentclass[A_4paper,11pt, final]{article}
\pdfoutput=1
\usepackage[utf8]{inputenc}
\usepackage{jheppub}
\usepackage{amsmath}
\usepackage{float}

\newcommand*\diff{\mathrm{d}}

\usepackage{xspace}
\newcommand*{\ie}{i.e., }
\newcommand*{\eg}{e.g., }
\newcommand*{\fig}{fig.\@\xspace}
\newcommand*{\eq}{eq.\@\xspace}
\newcommand*{\eqs}{eqs.\@\xspace}

\title{Coupling Metric-Affine Gravity\newline to a Higgs-Like Scalar Field}

\author[a]{Claire Rigouzzo,}
\author[a]{Sebastian Zell}

\affiliation[a]{Institue of Physics, Laboratory for Particle Physics and Cosmology,\\
	\'Ecole Polytechnique F\'ed\'erale de Lausanne, CH-1015 Lausanne, Switzerland}

\emailAdd{claire.rigouzzo.physics@hotmail.com}
\emailAdd{sebastian.zell@epfl.ch}

\abstract{
	General Relativity (GR) exists in different formulations. They are equivalent in pure gravity but generically lead to distinct predictions once matter is included. After a brief overview of various versions of GR, we focus on metric-affine gravity, which avoids any assumption about the vanishing of curvature, torsion or non-metricity. We use it to construct an action of a scalar field coupled non-minimally to gravity. It encompasses as special cases numerous previously studied models. Eliminating non-propagating degrees of freedom, we derive an equivalent theory in the metric formulation of GR. Finally, we give a brief outlook to implications for Higgs inflation. }

\date{}

\makeatletter
\gdef\@fpheader{\phantom{text}}
\makeatother

\begin{document}
	
	\maketitle
	
	\section{Introduction}
	\label{sec:intro}
	
	\subsection{The ambiguities of General Relativity}
	Einstein's theory of General Relativity (GR) describes gravity in terms of the geometry of spacetime. In its original version \cite{Einstein:1915}, it is solely based on curvature -- the rotation of vectors along closed curves. Correspondingly, the metric $g_{\mu\nu}$ is the unique fundamental variable, \ie there are only equations of motion for  $g_{\mu\nu}$. The affine connection is determined a priori as a function of $g_{\mu\nu}$. Its Christoffel symbols are defined by the conditions $\mathring{\Gamma}^\alpha_{~~\beta \gamma}= \mathring{\Gamma}^\alpha_{~~\gamma \beta }$ and $\mathring{\nabla}_\alpha g_{\mu\nu}=0$, which leads to the unique Levi-Civita connection. One can call this approach the metric formulation of GR.
	
	It was soon realized that there is another possibility: One can treat the metric $g_{\mu\nu}$ and the Christoffel symbols $\Gamma^\alpha_{~~\beta \gamma}$ as independent and regard both of them as fundamental variables \cite{Weyl:1918, Palatini:1919,Weyl:1922, Eddington:1923, Cartan:1922, Cartan:1923, Cartan:1924, Cartan:1925,Einstein:1925, Einstein:1928, Einstein:19282}.\footnote
	{A historical discussion can be found in \cite{Ferraris:1981}. Translations of \cite{Palatini:1919, Cartan:1922} are provided in \cite{Hojman:1980, Kerlick:1980}, and the works \cite{Einstein:1925,Einstein:1928, Einstein:19282} are translated in \cite{Unzicker2005}.}
	In this case, the Christoffel symbols $\Gamma^\alpha_{~~\beta \gamma}$ can deviate from the Levi-Civita connection $\mathring{\Gamma}^\alpha_{~~\beta \gamma}$ and are determined by their own equations of motions. This leads to the emergence of two additional geometrical concepts. The first one, proposed by Cartan \cite{Cartan:1922, Cartan:1923, Cartan:1924, Cartan:1925}, is torsion $T^\alpha_{~~\beta \gamma} \propto \Gamma^\alpha_{~~\beta \gamma} - \Gamma^\alpha_{~~\gamma\beta }$, which corresponds to the non-closure of infinitesimal parallelograms. The second one was put forward by Weyl \cite{Weyl:1918, Weyl:1922} and consists in non-metricity $Q_{\alpha\mu\nu}\propto \nabla_\alpha g_{\mu\nu}$.
	It causes the non-conservation of vector norms in parallel transport.
	
	The metric formulation of GR is based on the assumption that both torsion and non-metricity vanish, \ie that gravity is solely characterized by the curvature of spacetime. 
	It is possible, however, to relax these conditions.  
	Already in 1918, Weyl proposed a theory that features non-metricity in addition to curvature \cite{Weyl:1918} (see also \cite{Weyl:1922, Eddington:1923}).
	If, in contrast, solely torsion is included on top of curvature, this leads to the Einstein-Cartan formulation of gravity \cite{Cartan:1922, Cartan:1923, Cartan:1924, Cartan:1925, Einstein:1928, Einstein:19282}. If all three geometric properties curvature, torsion and non-metricity are included, one obtains a general metric-affine theory of gravity \cite{Hehl:1976kt, Hehl:1976kv, Hehl:1976my, Hehl:1977fj}.
	The list of possible formulations does not end here. For example, one can consider different teleparallel equivalents of GR \cite{Einstein:1928, Einstein:19282, Moller:1961, Pellegrini:1963, Hayashi:1967se,Cho:1975dh, Hayashi:1979qx, Nester:1998mp, BeltranJimenez:2019odq}, in which curvature is assumed to vanish, or purely affine gravity \cite{Eddington:1923,Einstein:1925,Schroedinger:1950, Kijowski:1978}, where $\Gamma^\alpha_{~~\beta \gamma}$ is the only dynamical field.
	
	At first sight, these various versions of GR appear to be very different. However, all of them are fully equivalent to the metric variant as long as no other fields are coupled to gravity and the action of the theory is chosen to be sufficiently simple. In metric-affine formulations, which encompass Weyl and Einstein-Cartan gravity as special cases, this can \eg be achieved with the usual Einstein-Hilbert action $\int \diff^4 x \sqrt{-g}\, R$, where $R$ is the Ricci scalar. Then the equivalence to metric GR comes about as follows: If there is no matter, the equations of motion for $\Gamma^\alpha_{~~\beta \gamma}$ determine that torsion and non-metricity vanish. This means that the Levi-Civita connection emerges dynamically as their solution.
	\textit{Thus, the different formulations are indistinguishable in a theory of pure gravity, \ie they represent an inherent ambiguity of GR.}
	
	There are conceptual advantages of gravitational theories in which $g_{\mu\nu}$ and $\Gamma^\alpha_{~~\beta \gamma}$ are treated as independent. First, boundary terms can be defined without any need for an infinite counterterm \cite{Ashtekar:2008jw}. Secondly, Einstein-Cartan gravity can be derived as a gauge theory of the Poincaré group \cite{Utiyama:1956sy, Sciama:1962, Kibble:1961ba}, which puts gravity on the same footing as the other fields of the Standard Model. Thirdly, it may be regarded as more aesthetical to obtain the Levi-Civita connection not because of an a priori assumption about the vanishing of any of the geometrical properties but as a result of extremizing an action. Nevertheless, none of these arguments constitute an irrefutable reason to prefer one or the other formulation. 
	
	\subsection{Metric-affine gravity and special cases}
	Among the possible formulations of GR, metric-affine gravity stands out because it relies on a minimal number of assumptions. None of the three geometric properties -- curvature, torsion and non-metricity -- are assumed to vanish, and instead all of them are fixed dynamically by their equations of motion. Moreover, metric-affine gravity encompasses the Weyl, Einstein-Cartan and metric versions of GR as special cases. Therefore, we shall focus on the metric-affine formulation in the following.
	
	As with all formulations, the equivalence of the metric-affine and metric versions of GR is generically broken once gravity is coupled to matter. This happens in two ways. On the one hand, it is possible that matter fields source torsion and/or non-metricity even when they are minimally coupled to gravity. For example, such a phenomenon occurs for fermions, but the resulting effects are suppressed by powers of the Planck mass $M_P$ \cite{Kibble:1961ba, Rodichev:1961}. On the other hand, one can extend the Einstein-Hilbert action by additional terms composed of torsion, non-metricity and possibly matter fields \cite{Hojman:1980kv, Nelson:1980ph,Nieh:1981ww, Percacci:1990wy, Castellani:1991et,Hehl:1994ue, Holst:1995pc,Obukhov:1996pf, Obukhov:1997zd,Shapiro:2001rz, Freidel:2005sn, Alexandrov:2008iy,Diakonov:2011fs, Magueijo:2012ug, Pagani:2015ema,Rasanen:2018ihz, Shimada:2018lnm}. Such contributions come with a priori undetermined coupling constants. If they are sufficiently big, resulting effects can be visible already far below the Planck scale. 
	In the above discussion, we did not mention terms with quadratic or higher powers of curvature since they generically lead to additional propagating degrees of freedom and therefore break the equivalence to GR even in the absence of matter \cite{Stelle:1977ry,Neville:1978bk,Neville:1979rb,Sezgin:1979zf,Hayashi:1979wj, Hayashi:1980qp}. Such models, which cannot be regarded as different formulations but correspond to modifications of gravity, will not be considered in the following.
	
	A remark is in order concerning naming. Sometimes the term Einstein-Cartan gravity is reserved for models with a minimal action, in which no contributions of torsion are present in addition to the Einstein-Hilbert term (see \eg \cite{Hehl:1980, Blagojevic:2002, Blagojevic:2003cg, Obukhov:2018bmf}).\footnote
	{In this case, the name Poincaré gauge theory is employed for gravitational models that feature torsion and a non-minimal action. In other cases, however, one only uses the term Poincaré gauge theory  when additional propagating degrees of freedom due to torsion are present (see \eg \cite{Hehl:1994ue}).}
	In contrast, Einstein-Cartan gravity will also denote torsionful theories with an extended action in the present paper. Analogously, we will use the term Weyl gravity for versions of GR with non-metricity, whether or not their action is minimal.\footnote
	{Weyl's original goal was to unify gravity and electromagnetism -- correspondingly, what we denote by Weyl gravity differs from the theory proposed in \cite{Weyl:1918}.}
	Finally, it remains to define what we mean by Palatini gravity. As has become convention, we will use this name for models with non-metricity, in which the purely gravitational part is minimal and only consists of the Ricci scalar.\footnote
	{It is interesting to note that non-metricity does not appear explicitly in Palatini's original work \cite{Palatini:1919} (see discussion in \cite{Ferraris:1981}).}
	This makes the Palatini version of GR a special case of Weyl gravity. As it turns out, choosing a minimal action in Einstein-Cartan gravity also leads to equivalence with the Palatini case.
	
	As a particular consequence of the equivalence among the different versions of GR, their particle spectra are identical and only consist of the two polarizations of the massless graviton. In a broad class of models, this is still the case in the presence of matter fields, \ie torsion and non-metricity are not dynamical but fully determined by algebraic equations in terms of the other fields. Therefore, it is possible to solve for $T^\alpha_{~\beta \gamma}$  and $Q_{\alpha\mu\nu}$. After plugging the results back into the original action, one obtains an equivalent theory in the metric formulation of gravity. \textit{In it the effects of torsion and non-metricity are replaced by a specific set of higher-dimensional operators in the matter sector}. Of course, one could have added such higher-dimensional terms from the very beginning, but then an effective field theory approach would have dictated to include all possible operators. In other words, allowing for generic gravitational geometries that feature torsion and non-metricity provides selection rules for singling out specific higher-dimensional operators in the matter sector.
	
	In the Einstein-Cartan formulation, \ie only considering torsion while still excluding non-metricity, various choices of matter fields and terms in the action have been considered and corresponding equivalent metric theories have been derived \cite{Nelson:1980ph, Castellani:1991et, Perez:2005pm, Freidel:2005sn,Alexandrov:2008iy,Taveras:2008yf, Torres-Gomez:2008hac, Calcagni:2009xz,Mercuri:2009zi, Diakonov:2011fs, Magueijo:2012ug,Langvik:2020nrs, Shaposhnikov:2020frq}. So far, the most complete study, taking into account all fields of the Standard Model, has been performed in \cite{Karananas:2021zkl}, which encompasses all previously cited papers as special cases. Additionally, criteria were developed and employed in \cite{Karananas:2021zkl} for systematically constructing an action of matter coupled to gravity. Their goal was to avoid making assumptions about the exclusion of possible terms, while still ensuring that the resulting theory is equivalent to metric GR in the absence of matter. This was achieved by only allowing contributions that are at most quadratic in torsion (or non-metricity) and at most linear in curvature. We note, however, that there are certain models with higher power of curvature which do not feature additional propagating degrees of freedom (see \cite{Karananas:2021zkl} for an example and a corresponding discussion). Therefore, the criteria of \cite{Karananas:2021zkl} are sufficient but not necessary for the absence of additional propagating degrees of freedom. Phenomenological implications of including such terms with higher powers of curvature, which do not bring about new particles, have been explored in \cite{Enckell:2018hmo, Antoniadis:2018ywb, Tenkanen:2019jiq, Gialamas:2019nly, Antoniadis:2019jnz, Lloyd-Stubbs:2020pvx, Antoniadis:2020dfq, Das:2020kff, Gialamas:2020snr, Dimopoulos:2020pas, Karam:2021sno, Lykkas:2021vax, Gialamas:2021enw,  Annala:2021zdt, Dioguardi:2021fmr}.
	
	Investigations in metric-affine gravity, where both torsion and non-metricity are present in addition to curvature, were mostly performed with a different approach, in which matter fields are not specified and no equivalent metric theory is derived.
	For example, general actions featuring all three of these geometric properties were proposed in \cite{Percacci:1990wy, Obukhov:1996pf, Obukhov:1997zd}, where only parity-even terms were taken into account. In this case, solutions for torsion and non-metricity in terms of the energy-momentum- and hypermomentum-tensors were obtained in \cite{Obukhov:1997zd}. 
	An action that also contains generic parity-odd terms was constructed in \cite{Pagani:2015ema}. 
	Based on \cite{Iosifidis:2021tvx}, solutions for torsion and non-metricity in terms of the energy-momentum- and hypermomentum-tensors were derived in this model in \cite{Iosifidis:2021bad}. An explicit computation of the equivalent metric theory was only performed in \cite{Rasanen:2018ihz}, where theories of a scalar field coupled to gravity in the metric-affine formulation were studied, but solely a specific subset of possible contributions due to torsion and non-metricity was included. A similar investigation with a simpler choice of action was performed in \cite{Shimada:2018lnm}.\footnote
	{Additionally, recent computations of an equivalent metric theory in a metric-affine model that includes fermions can be found in \cite{Delhom:2020gfv,  Delhom:2022xfo}.}

	The different formulations of GR have manifold cosmological implications. An incomplete list of relevant works includes \cite{Freidel:2005sn,Shie:2008ms, Taveras:2008yf, Torres-Gomez:2008hac,Chen:2009at,Baekler:2010fr,Poplawski:2011xf, Diakonov:2011fs, Khriplovich:2012xg, Magueijo:2012ug, Khriplovich:2013tqa, Kranas:2018jdc,Zhang:2019mhd,Zhang:2019xek,Saridakis:2019qwt,Barman:2019mlj,Aoki:2020zqm,Langvik:2020nrs,Shaposhnikov:2020gts,Shaposhnikov:2020aen, Iosifidis:2021iuw, Benisty:2021sul,Piani:2022gon} for the case of Einstein-Cartan gravity and  \cite{Obukhov:1997zd,Minkevich:1998cv,Puetzfeld:2001hk,Babourova:2002fn,Shimada:2018lnm, Iosifidis:2020zzp,Mikura:2020qhc, Iosifidis:2020upr,Kubota:2020ehu,Mikura:2021ldx,Iosifidis:2021kqo, Iosifidis:2021fnq} for generic metric-affine theories (see \cite{Puetzfeld:2004yg} for a guide to the literature up to 2004).\footnote
	{We remark that in some of the cited works additional propagating degrees of freedom are included in the gravity sector on top of the massless graviton.}
	The existence and characteristics of these effects due to torsion and non-metricity depend on the choice of gravitational formulation. In a theory of pure gravity, however, all versions of GR are equivalent and therefore on the same footing. This can spoil the uniqueness of observable predictions and makes it necessary to systematically explore phenomenological consequences of the different formulations of GR. In this way, we can hope to ultimately distinguish between them by observations and experiments. It is important to reiterate that we do not discuss modifications of gravity, but solely explore the consequences of the ambiguities that are inevitably contained in GR. If we do not want results to depend on potentially unjustified assumptions about the formulation of gravity, we have no choice but to investigate all of them.
	
	The goal of the present paper is to contribute to this program. We shall consider a scalar field coupled to gravity in a generic metric-affine formulation of GR, which includes both torsion and non-metricity in addition to curvature. First, we will construct a corresponding action by employing the criteria developed in \cite{Karananas:2021zkl}. Subsequently, we will solve for torsion and non-metricity and plug the results back in the action. In this way, we obtain an equivalent metric theory with specific higher-order operators for the scalar field. Our investigation aims at unifying several of the investigations described above. First, we generalize the scalar part of \cite{Karananas:2021zkl} by including non-metricity in addition to torsion. Secondly, we develop further \cite{Iosifidis:2021bad} by making the matter sector explicit -- using a scalar field as example --  and then deriving the equivalent metric theory.
	Finally, our work generalizes the paper \cite{Rasanen:2018ihz}, where only a subset of terms were included in the study of a scalar field non-minimally coupled to metric-affine gravity. Throughout, our analysis will be classical.\footnote{It would be very interesting to investigate if the various formulations of GR have implications for different approaches to quantum gravity, \eg in the contexts of asymptotic safety \cite{Weinberg:1980, Reuter:1996cp, Berges:2000ew}, loop quantum gravity \cite{Ashtekar:1986yd, BarberoG:1994eia, Thiemann:2001gmi}, the swampland program \cite{Vafa:2005ui,Obied:2018sgi, Palti:2019pca} or quantum breaking \cite{Dvali:2013eja, Dvali:2014gua, Dvali:2017eba}.}
	
	\subsection{Connection to Higgs inflation}
	In order to deal with the ambiguities due to the different formulations of GR, a possible approach is to exclude any large coupling constants in the action. In such a case, effects that are sensitive to the presence of torsion and/or non-metricity are suppressed by powers of $M_P$ and generically of limited phenomenological relevance. However, it is not always possible to adopt such an attitude. 
	
	A famous example in which it fails is the proposal that the Higgs boson of the Standard Model caused a period of exponential expansion in the early Universe \cite{Bezrukov:2007ep}. This idea of Higgs inflation stands out among inflationary scenarios since it does not require the introduction of any propagating degrees of freedom beyond those that are already present in the SM and GR. Therefore, it fits well the fact that so far no such additional particles have been detected experimentally. Moreover, the predictions of Higgs inflation as derived in \cite{Bezrukov:2007ep} are in excellent agreement with recent observations of the cosmic microwave background \cite{Planck:2018jri, BICEP:2021xfz}.
	
	However, the scenario of Higgs inflation is only phenomenologically viable if a large coupling constant is introduced in the action -- in the original proposal, which employed the metric formulation of GR, this was a non-minimal coupling of the Higgs field and the Ricci scalar \cite{Bezrukov:2007ep}. But beyond the special case of metric GR, many more analogous terms exist for coupling the Higgs field non-minimally to gravity. As a large coupling constant is required in any case, there is no reason to exclude other large parameters in the action. Correspondingly, the predictions of Higgs inflation strongly depend on the choice of gravitational formulation and terms in the action \cite{Bauer:2008zj,Rasanen:2018ihz, Raatikainen:2019qey, Langvik:2020nrs, Shaposhnikov:2020gts}.\footnote
	{Moreover, studies of inflation driven by a non-minimally coupled scalar field were performed in purely affine and teleparallel formulations \cite{Azri:2017uor, Jarv:2021ehj}.}
	So far only specific special cases have been analyzed and a systematic study of Higgs inflation in different versions of GR remains to be completed.  By employing a generic metric-affine formulation, which encompasses as special cases both metric gravity and the formulations used in \cite{Bauer:2008zj,Rasanen:2018ihz, Raatikainen:2019qey, Langvik:2020nrs, Shaposhnikov:2020gts}, we intend to lay the groundwork for such an investigation.
	
	The outline of the paper is as follows. Section \ref{sec:Dev-Riemannian} is devoted to geometric preliminaries and a more detailed review of possible formulations of GR. Additionally, we will introduce the criteria developed in \cite{Karananas:2021zkl} for constructing an action of matter coupled to gravity. In section \ref{sec:Gen-Action}, we present our theory of a scalar field coupled to GR in the metric-affine formulation. We first solve for torsion and non-metricity and derive the equivalent metric theory. Subsequently, we show how the results of \cite{Rasanen:2018ihz}, \cite{Shimada:2018lnm} and \cite{Karananas:2021zkl} are reproduced as special cases. In section \ref{sec:HiggsInflation}, we give a brief outlook of implications for Higgs inflation and we conclude in section \ref{sec:conclusion}. Appendix \ref{app:formulas} contains a few useful formulas, in appendix \ref{app:parallel_transport} we show how parallel transport along a closed curve is affected by torsion and non-metricity and appendix \ref{app:independence_terms} discusses linear dependence of different torsion-contributions.
	
	\textbf{Remark}. Preliminary results of the present investigation already appeared in the master thesis by one of us \cite{Rigouzzo:2021}.
	
	\textbf{Conventions}. We work in natural units $M_P=\hbar=c=1$, where $M_P$ is the reduced Planck mass, and use the metric signature $(-1,+1,+1,+1)$.  The covariant derivative of a vector $A^\nu$ is defined as
	\begin{equation} \label{covariantDerivative}
		\nabla_\mu A^\nu= \partial_\mu A^\nu + \Gamma^\nu~_{\mu \alpha}A^\alpha \;,
	\end{equation}
	\ie the summation is done on the last index of the Christoffel symbol. Square brackets denote antisymmetrization, $T_{[\mu \nu]} \equiv \frac{1}{2} (T_{\mu\nu} - T_{\nu\mu})$, and round brackets indicate symmetrization, $T_{(\mu \nu)} \equiv \frac{1}{2} (T_{\mu\nu} + T_{\nu\mu})$.

	\section{Curvature, torsion and non-metricity}
	\label{sec:Dev-Riemannian}
	
	\subsection{Geometric picture}
	In order to make our presentation self-contained, we shall begin by reviewing textbook knowledge about curvature, torsion and non-metricity. The reader familiar with this material is invited to proceed to subsection \ref{ssec:decomposition}. More details about the subsequent discussions can be found in \cite{Misner:1973prb,Schutz:1980, Carroll:2004,Heisenberg:2018vsk}.
	
	A differentiable manifold is described by two a priori independent quantities: the metric and the affine connection. The metric $g_{\mu\nu}$ defines distances in the manifold while the connection -- via the corresponding Christoffel symbols $\Gamma^\alpha_{~\beta \gamma}$ -- determines how parallel transport relates the tangent spaces at different points. A vector $\xi^\alpha$ that is parallel transported along a curve $\gamma(s)$ satisfies the following equation:
	\begin{equation}
		\frac{\diff\xi ^\alpha}{\diff s} = -\Gamma^\alpha_{~\beta \gamma}\frac{\diff x^\beta}{\diff s}\xi^\gamma \;,
		\label{parallel_transport}
	\end{equation}
	where $s$ is an affine parameter. If $g_{\mu\nu}$ and  $\Gamma^\alpha_{~\beta \gamma}$ are regarded as independent fundamental fields, then the Christoffel symbols $\Gamma^\alpha_{~\beta \gamma}$ encode three distinct geometric properties.
	
\begin{figure}[h]
		\hspace{0.3\linewidth}
		\includegraphics[width=0.4\linewidth]{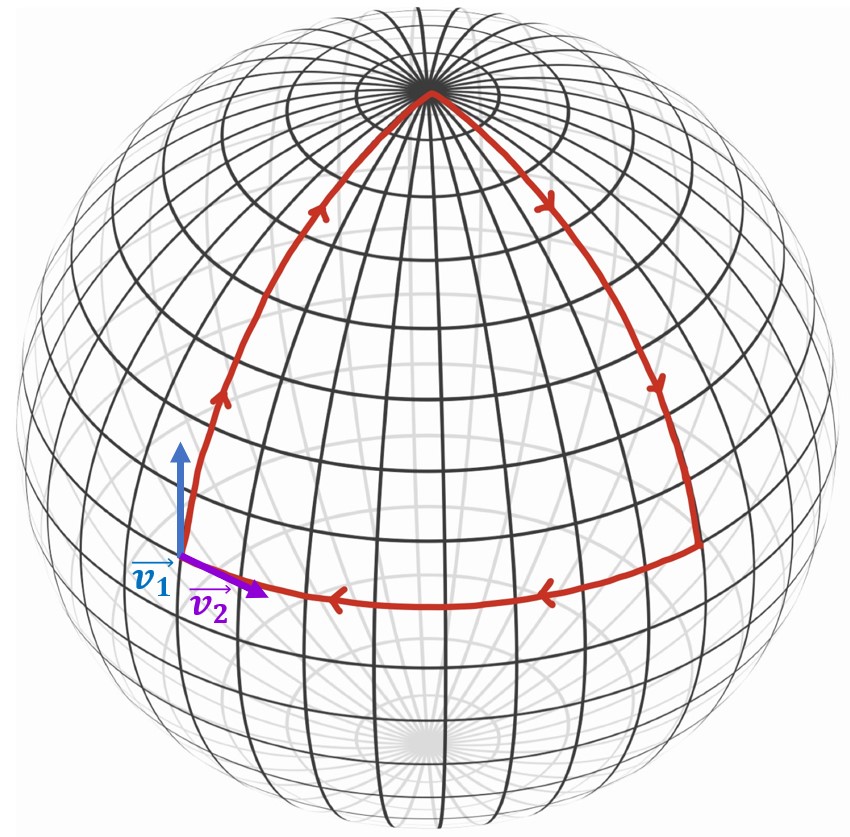}
		\caption{Representation of the effect of curvature. By parallel transporting the vector $\vec{v_1}$ along a closed path (in red), we obtain the vector $v_2$. The initial and the parallel transported vector do not coincide due to curvature.}
		\label{schema_curvature}
	\end{figure}

	\paragraph{Curvature}  
	The first one is curvature. It describes how parallel transport modifies the orientation of a vector, as illustrated in \fig \ref{schema_curvature}. For an infinitesimally small closed curve, the change of a vector $\xi^\alpha$ parallel transported along it is determined by the Christoffel symbols $\Gamma^\alpha_{~\beta \gamma}$ (see derivation in appendix \ref{app:parallel_transport}),
	
	\begin{equation}
		\Delta \xi^\alpha= \frac{1}{2}  \oint \diff \tau \frac{\diff x^\beta}{\diff \tau} x^\nu \xi^\gamma (0) 	R^\alpha_{\ \gamma\beta\nu}(0) \;,
		\label{curvature_change}
	\end{equation}
	where the Riemann tensor emerged:
	\begin{equation}
		R_{~\sigma \mu \nu}^{\rho}=\partial_{\mu} \Gamma_{~\nu \sigma}^{\rho}-\partial_{\nu} \Gamma_{~\mu \sigma}^{\rho}+\Gamma_{~\mu \lambda}^{\rho} \Gamma_{~\nu \sigma}^{\lambda}-\Gamma_{~\nu \lambda}^{\rho} \Gamma_{\mu \sigma}^{\lambda} \;.
		\label{curvature_general}
	\end{equation}
	As is evident, $R_{~\sigma \mu \nu}^{\rho}$ only is a function of $\Gamma^\alpha_{~\beta \gamma}$ and insensitive to the metric. Correspondingly, a framework in which $g_{\mu\nu}$ and $\Gamma^\alpha_{~\beta \gamma}$ are independent can be characterized as first-order formalism since the Riemann tensor only contains first derivatives.
	
\begin{figure}[h]
		\hspace{0.3\linewidth}
		\includegraphics[width=0.5\linewidth]{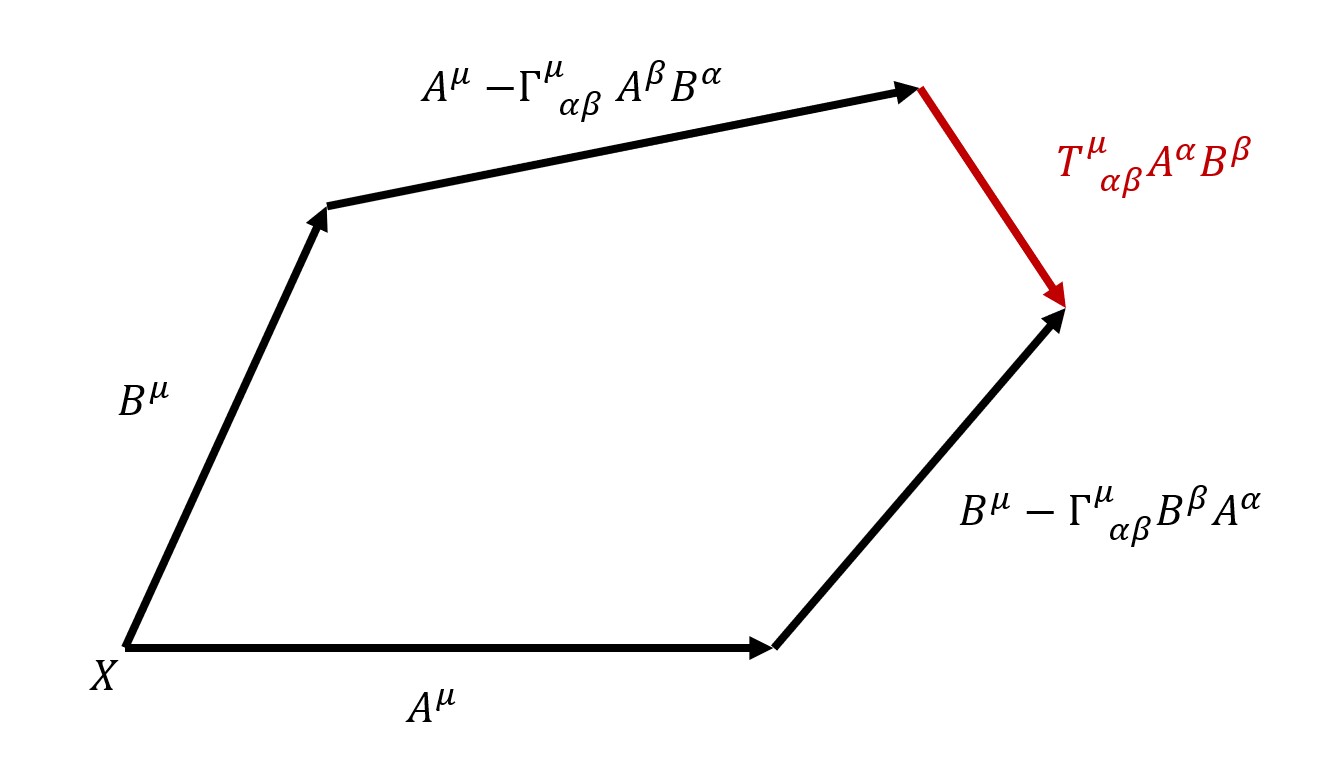}
		\caption{Representation of the effect of torsion. Two vectors $A^\mu$ and $B^\mu$ are parallel transported along each other. The non-closure of the resulting parallelogram, which is displayed in red, is proportional to torsion.}
		\label{schema_torsion}
	\end{figure}

	\paragraph{Torsion}
	The second geometric property is torsion \cite{Cartan:1922, Cartan:1923, Cartan:1924, Cartan:1925}. It is defined by
	\begin{equation} \label{torsionDefinition}
		T^\alpha_{~\beta \gamma} \equiv 2 \Gamma^\alpha_{~[\beta \gamma]}= \Gamma^\alpha_{~\beta \gamma}-\Gamma^\alpha_{~\gamma \beta} \;,
	\end{equation}
	\ie it emerges when the Christoffel symbols $\Gamma^\alpha_{~\beta \gamma}$ are not symmetric in the two lower indices.
	If torsion is present, then the parallelogram formed by the parallel transport of two vectors may not close, as represented in \fig \ref{schema_torsion}.
	Indeed if we consider two infinitesimal vectors $A^\mu (x^\nu)$ and $B^\mu(x^\nu)$ and we parallel transport them along each other, we obtain (see \eg also \cite{Yepez:2011bw}):  
	\begin{equation}
		\begin{split}
			A^{\mu}(x^\nu+B^\nu)= A^\mu (x^\nu) - \Gamma^\mu_{~\alpha \beta} A^\beta B^\alpha \;,\\
			B^\mu(x^\nu+A^\nu)=B^\mu(x^\nu) - \Gamma^\mu_{~\alpha \beta} B^\beta A^\alpha \;,
		\end{split} 
	\end{equation}
	where we used \eq \eqref{parallel_transport}.
	Hence the difference between the two transported vectors is : 
	\begin{equation}
		\begin{split}
			&A^{\mu}(x^\nu) + 	B^\mu(x^\nu+A^\nu) - \big(B^\mu(x^\nu) + 	A^{\mu}(x^\nu+B^\nu)\big)\\
			&=	\Gamma^\mu_{~\alpha \beta} A^\alpha B^\beta -\Gamma^\mu_{~\alpha \beta} B^\alpha A^\beta= 2 \Gamma^\mu_{~[\alpha \beta]} A^\alpha B^\beta = T^\mu_{~\alpha \beta}A^\alpha B^\beta \;,
		\end{split}
	\end{equation}
	\ie it is determined by the torsion tensor $T^\mu_{~\alpha \beta}$.

\begin{figure}[h]
		\includegraphics[width=1\linewidth]{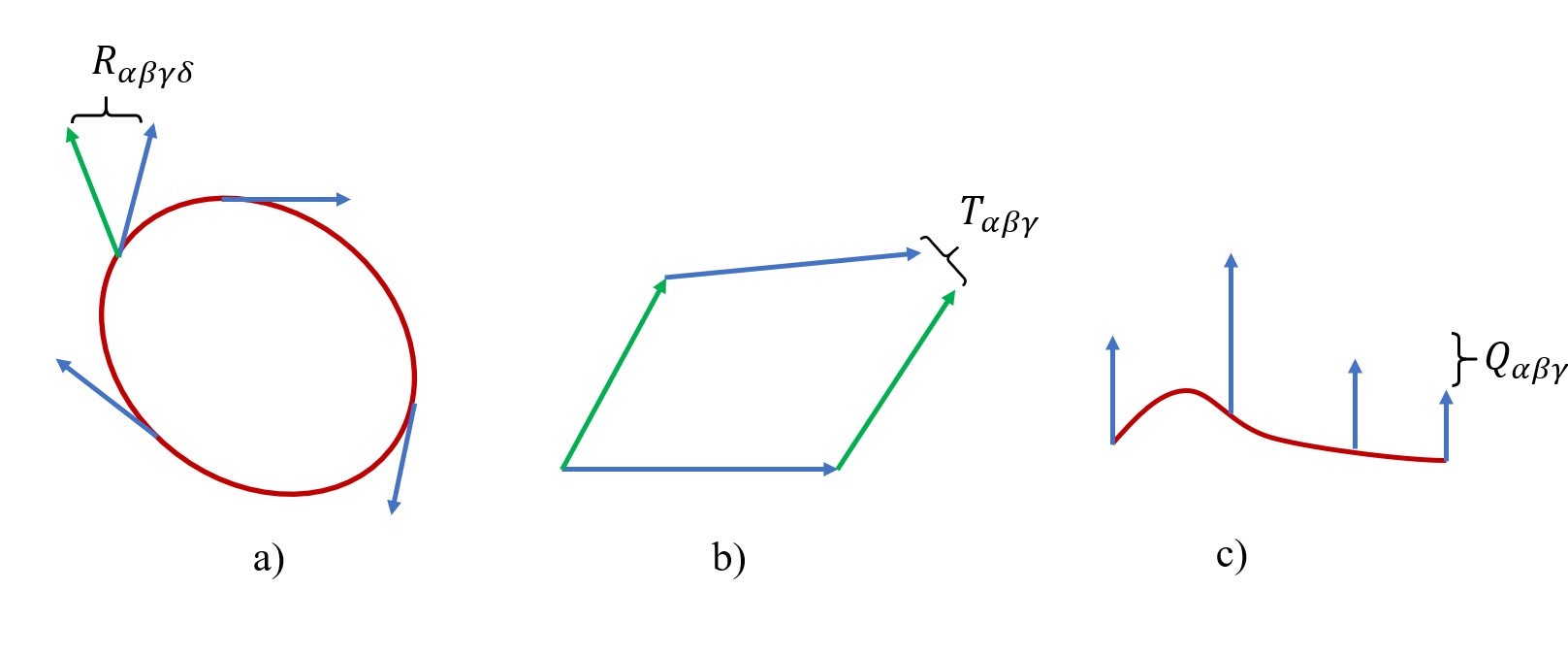}
		\caption{Schematic representation of the change of a vector under parallel transport due to the presence of: a) curvature b) torsion c) non-metricity. Figure inspired by \cite{BeltranJimenez:2019tjy}.}
		\label{schema}
	\end{figure}

	\paragraph{Non-metricity}
	Finally, the third geometric property is non-metricity. It emerges if the covariant derivative of the metric does not vanish and is defined by:
	\begin{equation}
		Q_{\gamma \alpha \beta} \equiv \nabla_{\gamma}g_{\alpha \beta} \;.
	\end{equation}
	In the presence of non-metricity, the norm of the vector may change under parallel transport.
	Indeed, we can consider the length of a vector $\xi^\alpha$ that is parallel transported:
	\begin{equation}
		\begin{split}
			\frac{\diff v^2}{\diff s}&=\frac{\diff (g_{\mu \nu} v^\mu v^\nu)}{\diff s} \\ &= \frac{\diff x^\alpha}{\diff s} \nabla_\alpha (g_{\mu \nu} v^\mu v^\nu) \\ &= \frac{\diff x^\alpha}{\diff s} (\nabla_\alpha g_{\mu \nu} v^\mu v^\nu+2 \underbrace{ \nabla_\alpha v^\mu}_{=0} g_{\mu \nu} v^\nu) \\&= \frac{\diff x^\alpha}{\diff s}  Q_{\alpha \mu \nu} v^\mu v^\nu \neq 0 \;.
		\end{split}
	\end{equation}
	To summarize, a schematic representation of curvature, torsion, and non-metricity is shown in \fig \ref{schema}.
	
	\paragraph{Special case: Riemannian geometry} As a special case, it is possible to consider a connection $\mathring{\Gamma}^\alpha_{~\beta \gamma}$ with vanishing torsion and non-metricity:
	\begin{equation} \label{LeviCivitaAssumptions}
		Q_{\mu \alpha \beta}=\mathring{\nabla}_\mu g_{\alpha \beta}=0 \;, \qquad T^\alpha_{\beta\gamma} = \mathring{\Gamma}^\alpha_{~\beta \gamma} - \mathring{\Gamma}^\alpha_{~ \gamma \beta}= 0 \;,
	\end{equation}
	where $\mathring{\nabla}$ is the covariant derivative associated with $\mathring{\Gamma}^\alpha_{~\beta \gamma}$.
	Once these conditions are imposed, the connection is uniquely determined as a function of the metric $g_{\mu \nu}$:
	\begin{equation}
		\mathring{\Gamma}^\alpha_{~\beta \gamma}=\frac{1}{2}g^{\alpha \mu}(\partial_\beta g_{\mu \gamma}+\partial_\gamma g_{\mu \beta}-\partial_\mu g_{\beta \gamma}) \;.
		\label{levicivita}
	\end{equation}
	The requirements \eqref{LeviCivitaAssumptions} lead to a Riemannian geometry and $\mathring{\Gamma}^\alpha_{~\beta \gamma}$ is the Levi-Civita connection. According to \eq \eqref{curvature_general}, the corresponding Riemann tensor reads:
	\begin{equation} \label{riemannLeviCivita}
		\mathring{R}_{~\sigma \mu \nu}^{\rho}=\partial_{\mu} \mathring{\Gamma}_{~\nu \sigma}^{\rho}-\partial_{\nu} \mathring{\Gamma}_{~\mu \sigma}^{\rho}+\mathring{\Gamma}_{~\mu \lambda}^{\rho} \mathring{\Gamma}_{~\nu \sigma}^{\lambda}-\mathring{\Gamma}_{~\nu \lambda}^{\rho} \mathring{\Gamma}_{\mu \sigma}^{\lambda} \;.
	\end{equation}
	Using the Levi-Civita connection $\mathring{\Gamma}^\alpha_{~\beta \gamma}$ leads to the metric formulation of GR. Since in this case second derivatives of the metric appear in \eq \eqref{riemannLeviCivita}, one can call it a second-order formalism. 
	
	It is worth noting that there is an asymmetry between curvature on the one side and torsion/non-metricity on the other side. Whereas curvature does not influence torsion/non-metricity, \eq \eqref{curvature_change} shows that torsion and non-metricity contribute to curvature (see also computation in appendix \ref{app:parallel_transport}). Correspondingly, an assumption about the absence of torsion and/or non-metricity, displayed in \eq \eqref{LeviCivitaAssumptions}, has different consequences than assuming that curvature vanishes. We will elaborate on this point in section \ref{ssec:possibleTheories}.

	\subsection{Decomposition of torsion and non-metricity}
	\label{ssec:decomposition}
	In full generality, we can decompose the connection into its Levi-Civita part and deviations from Riemannian geometry: 
	\begin{equation}
		\Gamma^\gamma_{~\alpha \beta}= \mathring{\Gamma}^\gamma_{~\alpha \beta}(g)+J^{\gamma}_{~\alpha \beta}(Q)+K^{\gamma}_{~\alpha \beta}(T) \;.
		\label{decomposition_connection}
	\end{equation}
	Here $\mathring{\Gamma}^\gamma_{\alpha \beta}(g)$ is the Levi-Civita connection, which only depends on the metric, $K^\gamma_{~\alpha \beta}(T)$ corresponds to the contorsion tensor depending on the torsion, and $J^\gamma_{~\alpha \beta}(Q)$ is the disformation tensor depending on the non-metricity. 
	Since contorsion $K^{\gamma}_{~\alpha \beta}$ is defined to be insensitive to non-metricity, it follows that $\nabla_\alpha g_{\mu \nu}\vert_{J^{\gamma}_{~\alpha \beta}=0}=0$. This condition determines contorsion as a function of torsion:
	\begin{equation}
		K_{\alpha \beta \gamma}=\frac{1}{2}(T_{\alpha \beta \gamma}+T_{\beta \alpha \gamma}+T_{\gamma \alpha \beta}) \;,
		\label{contorsion_torsion}
	\end{equation}
	where as usual $K_{\alpha \beta \gamma} \equiv g_{\alpha \sigma} K^{\sigma}_{~\beta \gamma}$. Similarly, we find the expression of disformation in terms of non-metricity by imposing that $\Gamma^\gamma_{~[\alpha \beta]}\vert_{J_{\alpha \beta \gamma}=0}=0$. This leads to:
	\begin{equation}
		J_{\alpha \mu \nu}=\frac{1}{2}(Q_{\alpha \mu \nu}-Q_{\nu \alpha \mu}-Q_{\mu \alpha \nu}) \;.
		\label{disformation}
	\end{equation}
	Note that \eq \eqref{contorsion_torsion} is sensitive to the convention \eqref{covariantDerivative} for the covariant derivative whereas \eq \eqref{disformation} is not. Contorsion is anti-symmetric in the first and last indices, $K_{\alpha\beta\gamma}=K_{[\alpha|\beta|\gamma ]}$, while disformation is symmetric in last two indices, $J_{\alpha \beta \gamma}=J_{\alpha (\beta \gamma)}$. 
	
	We can invert relations \eqref{contorsion_torsion} and \eqref{disformation} to express torsion in terms of contorsion and non-metricity in terms of disformation : 
	\begin{equation}
		\begin{split}
			Q_{\alpha \beta \gamma}=-2J_{(\beta| \alpha |\gamma)} \;, \qquad
			T_{\alpha \beta \gamma}=2K_{\alpha[\beta \gamma]}\;. 
		\end{split}
	\end{equation}
	This shows that contorsion (respectively disformation) and torsion (respectively non-metricity) encode the same information, because we can go from one to the other with a bijective transformation. Therefore, we can either view $\Gamma^\gamma_{~\alpha \beta}$ as fundamental field or $T^\gamma_{~\alpha \beta}$ and $Q^\gamma_{~\alpha \beta}$. Practically, this means that varying the action  with respect to $\Gamma^\gamma_{~\alpha \beta}$ is equivalent to a simultaneous variations with respect to $T^\gamma_{~\alpha \beta}$ and $Q^\gamma_{~\alpha \beta}$.
	
	We can further split torsion and non-metricity in vector- and tensor-parts. For torsion, irreducible representations are given by \cite{Hehl:1994ue,Obukhov:1997zd,Shapiro:2001rz}:
	\begin{align}
		&\text{the trace vector:\ } T^{\alpha}  =g_{\mu \nu}T^{\mu \alpha \nu} \;, \label{torsionTrace}\\
		&	\text{the pseudo trace axial vector:\ } \hat{T}^{\alpha}=\epsilon^{\alpha \beta \mu \nu}T_{\beta \mu \nu} \;, \label{torsionAxial}\\
		&\text{the pure tensor part:\ } t^{\alpha \beta \gamma}  \text{\ that satisfies\ } g_{\mu \nu}t^{\mu \alpha \nu}=0=\epsilon^{\alpha \beta \mu \nu}t_{\beta \mu \nu} \label{torsionTensor} \;.
	\end{align}
	Torsion can be expressed uniquely in terms of these irreducible pieces as:
	\begin{equation}
		T_{\alpha \beta \gamma }= -\frac{2}{3}g_{\alpha [\beta}T_{\gamma]}+\frac{1}{6}\epsilon_{\alpha \beta \gamma \nu}\hat{T}^\nu +t_{\alpha \beta \gamma} \;.
		\label{irrep_torsion}
	\end{equation}
	Similarly, we can split non-metricity into three pieces \cite{Hehl:1994ue,Obukhov:1997zd}: 
	\begin{align}
		&	\text{a first vector:\ } Q^\gamma=g_{\alpha \beta }Q^{\gamma \alpha \beta}\;, \label{nonMetricityVector1}\\
		&\text{a second vector:\ } \hat{Q}^\gamma=g_{\alpha \beta}Q^{\alpha \gamma \beta}\;, \label{nonMetricityVector2}\\
		&\text{the pure tensor part: } q^{\alpha \beta \gamma} \text{\ that satisfies\ } g_{\alpha \beta }q^{\gamma \alpha \beta}=0=g_{\alpha \beta}q^{\alpha \gamma \beta} \;. \label{nonMetricityTensor}
	\end{align}
	 Note that this decomposition does not correspond to irreducible representations since a fully symmetric tensor can still be separated from the pure tensor part $q^{\alpha \beta \gamma}$ \cite{Hehl:1994ue,Obukhov:1997zd}. In what follows, however, it will not be useful to further split $q^{\alpha \beta \gamma}$.
	Non-metricity can be expressed uniquely in terms of the components of \eqs \eqref{nonMetricityVector1} to \eqref{nonMetricityTensor}:
	\begin{equation}
		Q_{\alpha \beta \gamma}= \frac{1}{18}[g_{\beta \gamma}(5Q_{\alpha}-2\hat{Q}_{\alpha})+2g_{\alpha(\beta}(4\hat{Q}_{\gamma)}-Q_{\gamma)})]+ q_{\alpha \beta \gamma} \;.
		\label{irrep_metricity}
	\end{equation}
	As is evident from \eqref{torsionTrace} to \eqref{irrep_torsion}, the mapping of the full torsion tensor $T_{\alpha \beta \gamma }$ to the irreducible components $T^{\alpha}$, $\hat{T}^{\alpha}$ and $t^{\alpha \beta \gamma}$ is bijective. Eqs.\@\xspace \eqref{nonMetricityVector1} to \eqref{irrep_metricity} show that an analogous statement holds for the full non-metricity tensor $	Q_{\alpha \beta \gamma}$ and the contributions $Q^\gamma$, $\hat{Q}^\gamma$, and $q^{\alpha \beta \gamma}$. Since also the mapping between $T^\gamma_{~\alpha \beta}$ and $Q^\gamma_{~\alpha \beta}$ on the one hand and the full connection $\Gamma^\gamma_{~\alpha \beta}$  on the other hand is bijective, we conclude that a variation with respect to $\Gamma^\gamma_{~\alpha \beta}$ is equivalent to a simultaneous variations with respect to the $6$ tensors  $T^{\alpha}$, $\hat{T}^{\alpha}$, $t^{\alpha \beta \gamma}$, $Q^\gamma$, $\hat{Q}^\gamma$ and $q^{\alpha \beta \gamma}$.
	Finally, let us discuss the number of independent components in each irreducible piece.
	First, $T_{\alpha \beta \gamma}$ is antisymmetric in its last two indices, yielding $4\times 6=24$ independent components. Because $T^\alpha$ and $\hat{T}^\alpha$ are vectors, they can only have $4$ independent terms, and $t^{\alpha \beta \gamma}$ then carries the remaining $16$ independent components. For non-metricity the number is higher because $Q_{\alpha \beta \gamma}$ is symmetric in the last two indices, leading to $4\times 10=40$ independent components. Following the same argument, $Q^\alpha$ and $\hat{Q}^\alpha$ each carry $4$ independent components while $q^{\alpha \beta \gamma}$ carries $32$. Overall, the sum reproduces $64$ independent components of the initial affine connection $\Gamma^\alpha_{~\beta \gamma}$, in accordance with bijectivity.
	
	Using the decomposition \eqref{decomposition_connection} of the connection as well as formulas \eqref{Qdecomposition1} - \eqref{QTdecomposition} from appendix \ref{app:formulas}, we can split the Ricci scalar as follows (see also \cite{Langvik:2020nrs}):
	\begin{equation}
		\begin{split}
			R&=\mathring{R}+\mathring{\nabla}_\alpha (Q^\alpha - \hat{Q}^\alpha +2 T^\alpha) -\frac{2}{3}T_\alpha (T^\alpha+Q^\alpha-\hat{Q}^\alpha)  +\frac{1}{24}\hat{T}^\alpha \hat{T}_\alpha + \frac{1}{2} t^{\alpha \beta \gamma} t_{\alpha \beta \gamma } \\
			&-\frac{11}{72} Q_\alpha Q^\alpha+ \frac{1}{18}\hat{Q}_\alpha \hat{Q}^\alpha+ \frac{2}{9} Q_\alpha \hat{Q}^\alpha+\frac{1}{4}q_{\alpha \beta \gamma}(q^{\alpha \beta \gamma}-2 q^{\gamma \alpha \beta}) +t_{\alpha \beta \gamma } q^{\beta \alpha  \gamma},
		\end{split}
		\label{curvatureSplit}
	\end{equation}
	where $\mathring{R}=g_{\mu \nu}\mathring{R}^\alpha_{~\mu\alpha\nu}$ is the scalar curvature solely computed from the Levi-Civita connection $\mathring{\Gamma}^\alpha_{~~\beta \gamma}$, as derived from the Riemann tensor shown in \eq \eqref{riemannLeviCivita}. 
	
	The scalar curvature $R$ obeys an interesting property: it is invariant under projective transformation, defined by \cite{Schroedinger:1950,Trautman1973,Sandberg:1975db,Hehl:1976kv,Trautmann1976,Hehl:1978, Hehl:1981}:
	\begin{equation}
		\Gamma^\gamma_{~\alpha \beta} \rightarrow \Gamma^\gamma_{~\alpha \beta} + \delta^\gamma_\beta A_\alpha  \;,
		\label{projective}
	\end{equation}
	with $A_{\alpha}=A_{\alpha}(x)$ an arbitrary covariant vector field. Geometrically, \eq \eqref{projective} represents the most general transformation that changes the auto-parallel curves by a reparametrization of their affine parameter (see \cite{Iosifidis:2018zjj} for details). 
	Notably, most irreducible components are not invariant under \eq \eqref{projective}: 
	\begin{equation}
		T^\alpha \rightarrow  T^\alpha + 3 A^\alpha, \quad \hat{T}^\alpha \rightarrow \hat{T}^\alpha, \quad
		Q^\alpha \rightarrow Q^\alpha -8 A^\alpha, \quad	\hat{Q}^\alpha \rightarrow \hat{Q}^\alpha -2 A^\alpha \;,
	\end{equation}
	but the combination that enters into the scalar curvature given in \eq \eqref{curvatureSplit}, and correspondingly the Einstein-Hilbert action, are invariant. As long as an action remains unchanged under projective transformations, the connection $\Gamma^\gamma_{~\alpha \beta}$ cannot be uniquely determined by its equations of motion \cite{Schroedinger:1950,Trautman1973,Sandberg:1975db,Hehl:1976kv,Trautmann1976,Hehl:1978, Hehl:1981}. However, a general theory may not be invariant under the projective transformation, as will be discussed in section \ref{sec:Gen-Action}.
	
	\subsection{Classifying possible theories}
	\label{ssec:possibleTheories}
	We have seen that a generic geometry of spacetime can be characterized by the three properties: curvature, torsion and non-metricity. When devising a theory of gravity, one has to decide for each of these three concepts whether they should be included or excluded. Therefore, up to eight choices are available to us.  Clearly, excluding all non-trivial geometry leads to a Minkowski spacetime and the absence of gravity, which leaves us with seven possibilities. As we shall discuss, all seven indeed lead to viable formulations of gravity,  which are summarized in table \ref{diff_formalism}. 

	Expanding on the introduction, we shall discuss them in the following. In doing so, we will briefly sketch how some of their properties can be derived. Our goal is to convey to the reader a rough idea of the underlying calculations in a manner that is as concise as possible. Therefore, we leave out many details and equations in the present subsection \ref{ssec:possibleTheories} are only symbolic. Precise computations for the metric-affine formulation will be presented in section \ref{sec:Gen-Action}. For teleparallel theories we refer the reader to the references displayed subsequently.
	
	\begin{table}
		\centering
		\begin{tabular}{|p{0.36\linewidth}|c|c|c|p{0.36\linewidth}|}
			\hline
			Formulation of gravity & $R_{\alpha \beta \gamma \delta}$  & $T_{\alpha \beta \gamma}$ & $Q_{\alpha \beta \gamma}$  & Equivalent to metric GR for arbitrary coefficients of $T^2$, $QT$, $Q^2$ 	\\ \hline
			Metric-affine  \cite{Hehl:1976kt, Hehl:1976kv, Hehl:1976my, Hehl:1977fj}   &   &     &   &
			Yes   \\ \hline
			Einstein-Cartan \cite{Cartan:1922, Cartan:1923, Cartan:1924, Cartan:1925, Einstein:1928, Einstein:19282}  &   &  &  $=0$ &
			Yes  \\\hline
			Weyl \cite{Weyl:1918,Weyl:1922,Eddington:1923} &   & $=0$ & & Yes \\ \hline
			Metric \cite{Einstein:1915}  &     & $=0$    & $=0$  & (Not applicable) \\ \hline \hline
			Generic teleparallel \cite{BeltranJimenez:2019odq}    & $=0$   &    &  & No  \\ \hline
			Metric teleparallel \cite{Einstein:1928, Einstein:19282, Moller:1961, Pellegrini:1963, Hayashi:1967se,Cho:1975dh, Hayashi:1979qx}    & $=0$   &    & $=0$   & No \\ \hline
			Symmetric teleparallel \cite{Nester:1998mp}&  $=0$   & $=0$  &     & No  \\ \hline                                                                                                                                                                                                                                                                                                    
		\end{tabular}
		\caption{List of different formulations of gravity. Properties are summarized, with a display of the vanishing of tensorial quantities. As in the text, $T^2$ stands for an arbitrary invariant composed of torsion $T_{\alpha \beta \gamma}$, and analogous statements apply to $QT$ and $Q^2$.}
		\label{diff_formalism}.
	\end{table}
	
	\subsubsection{Theories with curvature}
	First, we shall discuss the four possible formulations that feature curvature. Clearly, excluding a priori both torsion and non-metricity results in the most commonly-used metric version of GR \cite{Einstein:1915}. In the absence of matter, its action is given by
	\begin{equation} \label{actionMetricGravitySymbolical}
		\mathcal{L}_{\text{metric}} \sim \mathring{R} \;,
	\end{equation}
	where  $\mathring{R}$ is the curvature determined by the Levi-Civita connection, as defined in \eq \eqref{riemannLeviCivita}.
	Next, we shall discuss the effect of including the other two geometric concepts. As reviewed in the introduction, adding non-metricity in addition to curvature leads to Weyl gravity \cite{Weyl:1918,Weyl:1922}, whereas a theory that features both torsion and curvature corresponds to the Einstein-Cartan formulation \cite{Cartan:1922, Cartan:1923, Cartan:1924, Cartan:1925, Einstein:1928, Einstein:19282}.  Including all three geometric properties -- curvature, torsion and non-metricity -- results in a general metric-affine theory of gravity \cite{Hehl:1976kt, Hehl:1976kv, Hehl:1976my, Hehl:1977fj}; see \cite{Hehl:1980,Hehl:1994ue,Blagojevic:2002, Blagojevic:2003cg, Obukhov:2018bmf} for reviews.
	
	Once torsion and/or non-metricity are included, the next question is what action one should use. An obvious choice is 
	\begin{equation} \label{actionAffineGravitySymbolicalpecific}
		\mathcal{L}_{\text{affine, specific}} \sim R \;,
	\end{equation} 
	where the corresponding Riemann tensor is defined in \eqref{curvature_general}. Such a model, in which the purely gravitational action only consists of the Ricci scalar, leads to the Palatini version of GR (see discussion in section \ref{sec:intro}). As derived in \eq \eqref{curvatureSplit}, we can split $R$ in a part $\mathring{R}$ that solely depends on the Levi-Civita connection $\mathring{\Gamma}^\alpha_{~~\beta \gamma}$ and quadratic invariants composed of torsion and/or non-metricity. We note that we can leave out contributions of the form $\mathring{\nabla} T^\alpha$ and $\mathring{\nabla} Q^\alpha$ since they only lead to boundary terms. Moreover, the quadratic contributions of torsion and non-metricity in \eq \eqref{actionAffineGravitySymbolicalpecific} have fixed coefficients (\eg $Q_\alpha Q^\alpha$ comes with a factor of $-11/72$; see \eq \eqref{curvatureSplit}). However, we can be more general and include quadratic invariants with arbitrary coefficients. We will give the precise form of the resulting action in section \ref{sec:Gen-Action} (see \eq \eqref{general_action_component}). For now, we shall content ourselves with briefly sketching the effect of including torsion and/or non-metricity. To this end, it suffices to write symbolically
	\begin{equation} \label{actionAffineGravitySymbolical}
		\mathcal{L}_{\text{affine}} \sim  \mathring{R} + c_{TT} T^2 + c_{QQ} Q^2 +  c_{TQ} TQ \;,
	\end{equation}
	where $T$ and $Q$ stand for any tensor linear in torsion and non-metricity, respectively. (For example, $T$ can represent $T^\alpha, \hat{T}^\alpha$ and $t^{\alpha\beta\gamma}$). Moreover, $c_{TT}$, $c_{QQ}$ and $c_{TQ}$ are arbitrary coefficients. Now we can determine  torsion and/or non-metricity by their equations of motion. For the action \eqref{actionAffineGravitySymbolical}, a solution is given by
	\begin{equation}
		T = 0 \;, \qquad Q =0 \;.
	\end{equation}
	Thus, the two additional geometric properties vanish dynamically in the absence of matter. This shows why in purely gravitational theories of the form \eqref{actionAffineGravitySymbolical}, the metric-affine formulation as well as its two special cases Einstein-Cartan and Weyl-gravity are equivalent to the most commonly-used metric version of GR. 
	
	Next we shall repeat the previous discussion in the presence of matter, where we use a scalar field $h$ as an example. Motivated by an analogy to the Higgs field of the Standard Model in unitary gauge, we shall assume that $h$ possesses a $Z_2$-symmetry $h \rightarrow -h$. Apart from this property, however, $h$ will represent in the present work a generic scalar field which can be different from the Higgs field. In the metric formulation, the action for coupling $h$ to gravity is given by
	\begin{equation} \label{actionMetricGravityMatterSymbolical}
		\mathcal{L}_{\text{metric}} \sim \mathring{R}  + \xi h^2 \mathring{R} +  \mathcal{L}_\text{m}\;,
	\end{equation}
	which reduces to \eq \eqref{actionMetricGravitySymbolical} in the absence of matter.
	Here $\xi$ parametrizes a non-minimal coupling of the scalar field to gravity and $\mathcal{L}_\text{m}$ can contain all operators in the matter sector that are independent of the Christoffel symbols. Once torsion and/or non-metricity are included, the generalization of \eq \eqref{actionAffineGravitySymbolicalpecific} leads to
	\begin{equation} 
		\mathcal{L}_{\text{affine, specific}} \sim R + \xi h^2 R +  \mathcal{L}_\text{m} \;. 
	\end{equation} 
	Again we can use \eq \eqref{curvatureSplit} to split $R$ in a Levi-Civita part $\mathring{R}$ and terms involving torsion and/or non-metricity. Due to the non-minimal coupling of $h$ to $R$, now also the terms of the form $\mathring{\nabla}_\alpha T^\alpha$ and $\mathring{\nabla}_\alpha Q^\alpha$ give a non-trivial contribution. As before, we replace specific by arbitrary coefficients, and so the generalization of \eq \eqref{actionAffineGravitySymbolical} in the presence of matter yields 
	\begin{equation} \label{actionAffineMatterSymbolical}
		\begin{split}
		\mathcal{L}_{\text{affine}} \sim &   \mathring{R} + \xi h^2 \mathring{R} +(c_{TT} + \tilde{c}_{TT} h^2) T^2 + (c_{QQ} + \tilde{c}_{QQ}h^2) Q^2 +  (c_{TQ} + \tilde{c}_{TQ} h^2) TQ\\
		& +\xi_T h^2 \mathring{\nabla} T  +  \xi_Q h^2 \mathring{\nabla} Q  + \mathcal{L}_\text{m}  \;.
		\end{split}
	\end{equation}
	It is important to note that the number of coefficients describing a non-minimal coupling of matter to gravity has significantly increased. Whereas only one such parameter exists in metric gravity ($\xi$), many more analogous contributions emerge in the metric-affine formulation. Once torsion and/or non-metricity are present, there is no reason to exclude the couplings of matter to gravity, which are all on the same footing as the single non-minimal coupling term in metric gravity.
	
	In \eq \eqref{actionAffineMatterSymbolical}, the equations of motion for $J$ and $Q$ yield a non-trivial result:
	\begin{equation} \label{solutionAffineMatterSymbolical}
		T \sim \partial h^2 \;, \qquad Q \sim \partial h^2 \;.
	\end{equation}
	The significance of this finding is twofold. First, it shows how torsion and/or non-metricity are sourced once appropriate couplings to matter, such as a scalar field, are added. Secondly, the solution \eqref{solutionAffineMatterSymbolical} is algebraic. Thus, torsion and non-metricity do not propagate, and also metric-affine gravity only features excitations of a massless graviton. As a particular consequence, we can plug the solution \eqref{solutionAffineMatterSymbolical} back into the action \eqref{actionAffineMatterSymbolical}:\footnote
	{We remark that \eqs \eqref{actionAffineMatterSymbolical}, \eqref{solutionAffineMatterSymbolical} and \eqref{equivalentMetricSymbolical} are symbolic versions of \eqs \eqref{general_action_component}, \eqref{solution_short} and \eqref{action_mid}, respectively, which will be derived in the subsequent section \ref{sec:Gen-Action}.}
	\begin{equation} \label{equivalentMetricSymbolical}
		\mathcal{L}_{\text{affine}} \sim \mathring{R} + \xi h^2 \mathring{R}  + f(h) \left(\partial h^2\right)^2 + \mathcal{L}_\text{m} \;,
	\end{equation}
	where $f(h)$ is a function of $h$ that is determined by the parameters appearing in the action \eqref{actionAffineMatterSymbolical}.
	We can call \eq \eqref{equivalentMetricSymbolical} the equivalent metric theory, in which the effects of torsion and/or non-metricity are replaced by specific additional operators in the matter sector.
	
\begin{figure}[H]
		\centering
		\includegraphics[width=0.66\linewidth]{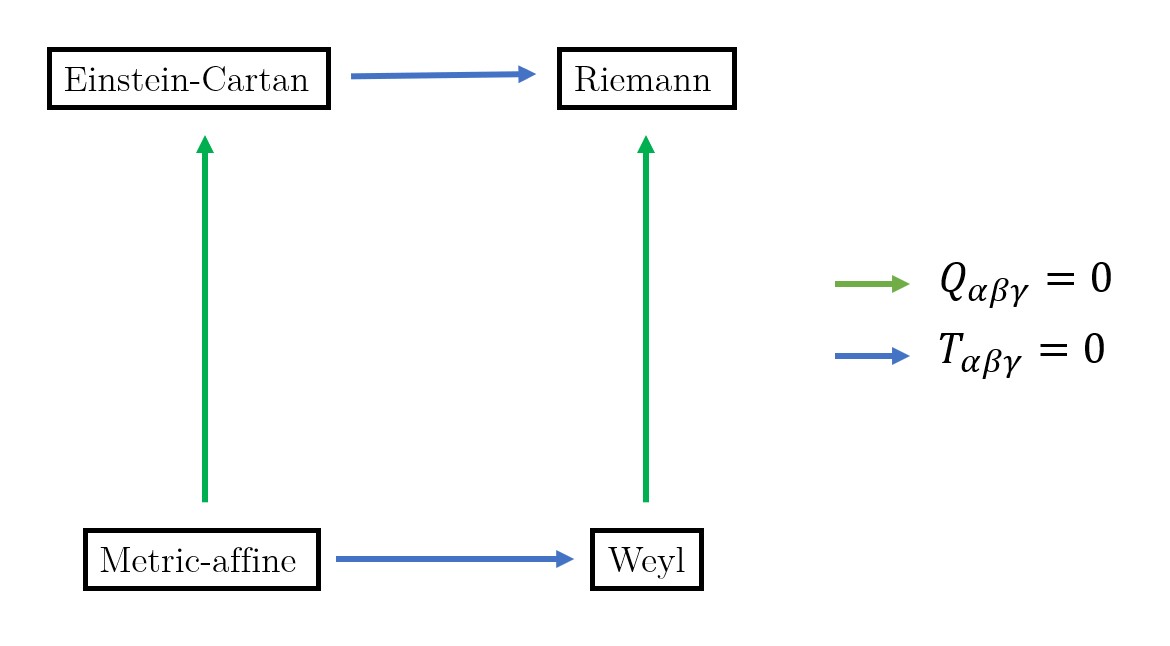}
		\caption{Diagram of relations between different formulations of gravity with curvature. Starting from the most general class of metric-affine theories of gravity in the lower, left corner; one can obtain specific limits by imposing that torsion or non-metricity vanish.}
		\label{schema_theories_curvature}
	\end{figure}

	In summary, we have outlined why in the presence of matter, the different formulations of gravity that feature curvature are no longer equivalent. Their difference can be reduced to specific additional interactions in the matter sector, which feature a number of a priori unknown coupling constants. Finally, we note that the limits of excluding torsion and/or non-metricity are smooth, \ie the following two procedures lead to the same result. On the one hand, one can assume a priori that torsion and/or non-metricity vanish. On the other hand, it is equivalent to put in the Lagrangian the coefficients of all terms involving $T$ and/or $Q$ to zero. To obtain the Einstein-Cartan formulation for example, one can simply set in \eq \eqref{actionAffineMatterSymbolical} all coefficients involving $Q$ to zero and arrive at an accordingly simplified equivalent metric theory \eqref{equivalentMetricSymbolical}.
	A summary of the relation between the different theories of gravity with curvature is given in \fig \ref{schema_theories_curvature}.
	
	\subsubsection{Teleparallel theories}
	Next we shall turn to three possible teleparallel formulations of GR, in which curvature is excluded; see \cite{Maluf:2013gaa,Aldrovandi:2013wha,Heisenberg:2018vsk,Krssak:2018ywd, Bahamonde:2021gfp} for reviews. First, a metric teleparallel theory was proposed, in which only torsion is present and non-metricity is assumed to vanish \cite{Einstein:1928, Einstein:19282, Moller:1961, Pellegrini:1963, Hayashi:1967se,Cho:1975dh, Hayashi:1979qx}.  Subsequently, a symmetric teleparallel formulation was developed that exclusively features non-metricity \cite{Nester:1998mp}. Only very recently, a general teleparallel theory was constructed that simultaneously contains both torsion and non-metricity \cite{BeltranJimenez:2019odq}. 
	The assumption of vanishing curvature has different implications than setting to zero torsion and/or non-metricity. Assuming that the latter two quantities vanish in a metric-affine formulation does not have any effects on the Levi-Civita curvature $\mathring{R}$. In contrast, this is not the case for the assumption of teleparallelism, as one can anticipate from  \eq \eqref{curvatureSplit}, which shows that curvature is the sum of a Levi-Civita part $\mathring{R}$ and contributions of torsion and non-metricity. Therefore, setting to zero curvature can constrain $\mathring{R}$ in terms of torsion and/or non-metricity. In the following, we shall briefly sketch how this comes about, and we refer the reader \eg to \cite{BeltranJimenez:2017tkd,BeltranJimenez:2018vdo,Heisenberg:2018vsk,  BeltranJimenez:2019odq,  Bahamonde:2021gfp} for more details.
	
	As in \cite{BeltranJimenez:2019odq}, we will include both torsion and non-metricity.\footnote
	{It is straightforward to leave out one of the two quantities, and analogous statements will apply.}
	First we consider the theory in the absence of matter, where we follow \cite{Bahamonde:2021gfp}. In analogy to \eq \eqref{actionAffineGravitySymbolicalpecific}, we start from the action 
	\begin{equation} \label{actionTeleparallelSymbolical}
		\mathcal{L}_{\text{teleparallel, specific}} \sim  -T^2   -Q^2   -TQ + \lambda \left(\mathring{R} +  T^2 +  Q^2 +  TQ + \mathring{\nabla} T + \mathring{\nabla} Q \right) \;,
	\end{equation}
	where the Lagrange multiplier  $\lambda = \lambda(x)$ enforces the constraint of vanishing curvature. Moreover, the coefficients of the quadratic invariants in torsion and/or non-metricity are fixed according to \eq \eqref{curvatureSplit}, up to a sign change in the first three terms. The motivation for this specific choice of parameters comes from ensuring equivalence to metric GR. Namely, we can plug the constraint $-T^2 -  Q^2 -  TQ = \mathring{R} + \mathring{\nabla} T + \mathring{\nabla} Q$ back in the action \eqref{actionTeleparallelSymbolical} to obtain
	\begin{equation}
		\mathcal{L}_{\text{teleparallel, specific}} \sim  \mathring{R} + \mathring{\nabla} T + \mathring{\nabla} Q \;.
	\end{equation}
	Up to a boundary term, this coincides with the result \eqref{actionMetricGravitySymbolical} of metric gravity. Therefore, \eq \eqref{actionTeleparallelSymbolical} is the action of the General Teleparallel Equivalent of GR (GTEGR) \cite{BeltranJimenez:2019odq}. Leaving out all terms involving non-metricity leads to the Metric Teleparallel Equivalent of GR \cite{Einstein:1928, Einstein:19282, Moller:1961, Pellegrini:1963, Hayashi:1967se,Cho:1975dh, Hayashi:1979qx}.\footnote
	{Since this theory was constructed first, it is often simply referred to as Teleparallel Equivalent of GR.}
	Correspondingly, omitting all contributions of torsion yields the Symmetric Teleparallel Equivalent of GR \cite{Nester:1998mp}. We shall not explicitly discuss how matter can be coupled to the different teleparallel equivalents of GR but only refer the reader to the literature. It was noted early on that in torsionful theories an issue can arise due to fermions \cite{Hayashi:1979qx} but that a consistent interaction with matter can be achieved with an appropriate choice of coupling prescription; see \cite{Hayashi:1979qx, deAndrade:1997cj, deAndrade:1997gka, deAndrade:2000kr, deAndrade:2001vx, Obukhov:2002tm,Maluf:2003fs, Mielke:2004gg,Mosna:2003rx, Obukhov:2004hv, Aldrovandi:2013wha} for studies excluding non-metricity and \cite{Adak:2008gd, BeltranJimenez:2017tkd, BeltranJimenez:2018vdo, Jimenez:2019woj, Delhom:2020hkb, BeltranJimenez:2020sih} for investigations without this restriction. We note that the Symmetric Teleparallel Equivalent of GR may evade some of the ambiguities caused by torsion \cite{BeltranJimenez:2017tkd, BeltranJimenez:2018vdo, Jimenez:2019woj, BeltranJimenez:2020sih}. 
	
	Finally, we shall give a brief outlook to generalizations of the Lagrangian \eqref{actionTeleparallelSymbolical}. First, one can attempt to choose arbitrary coefficients in \eq \eqref{actionTeleparallelSymbolical}, in analogy to our approach in the metric-affine case. For the case of vanishing non-metricity, this was already suggested in \cite{Hayashi:1979qx} under the name New GR and symbolically reads
	\begin{equation} \label{actionTeleparallelGenericSymbolical}
		\mathcal{L}_{\text{teleparallel, generic}} \sim c_{TT} T^2 +  \lambda \left(\mathring{R} +  T^2  + \mathring{\nabla} T  \right) \;,
	\end{equation}
	where we note that the parameters in the constraint remain fixed. However, issues were discovered in this model \cite{Kopczynski:1982,Kuhfuss:1986rb,Nester:1988, Cheng:1988zg}. Moreover, it generically contains additional propagating degrees of freedom \cite{Kuhfuss:1986rb}, and so it differs from metric GR already in the absence of matter and does not correspond to an equivalent formulation.\footnote
	{The fact that a derivative of torsion appears in the constraint $\mathring{R} +  T^2  + \mathring{\nabla} T = 0$ is already an indication that the teleparallel theory \eqref{actionTeleparallelGenericSymbolical} contains additional propagating degrees of freedom, unless specific values of the parameters are chosen; we refer the reader \eg to \cite{BeltranJimenez:2018vdo} for a detailed computation.}
	Analogous statements, namely that additional propagating degrees of freedom emerge for generic parameter choices, hold in the other teleparallel models. For a theory that only features non-metricity this question was studied in \cite{BeltranJimenez:2017tkd, BeltranJimenez:2018vdo}, where the term Newer GR was introduced, and a model with both torsion and non-metricity was investigated in \cite{BeltranJimenez:2019odq}. Thus, even though the geometry of a generic teleparallel theory is simpler than in the metric-affine case, its particle spectrum is more involved. Starting from a generic gravitational Lagrangian \eqref{actionTeleparallelGenericSymbolical}, equivalence with metric GR is only achieved for specific values of the coefficients. These parameter choices can arise as a result of symmetries \cite{BeltranJimenez:2017tkd, BeltranJimenez:2018vdo, Jimenez:2019woj, BeltranJimenez:2019odq}. Applications of teleparallel gravity to cosmology, \eg with respect to inflation and dark energy, can for example be found in \cite{Bengochea:2008gz, Linder:2010py,Myrzakulov:2010vz, Wu:2010xk,Chen:2010va, Bengochea:2010sg, Dent:2010nbw,Li:2011wu, Cai:2011tc,Sharif:2011bi, Hohmann:2017jao,BeltranJimenez:2017tkd,Jarv:2018bgs,BeltranJimenez:2018vdo,Dialektopoulos:2019mtr,BeltranJimenez:2019tme,Raatikainen:2019qey,DAmbrosio:2020nev, Bose:2020xdz,DAmbrosio:2021pnd,Solanki:2022rwu, Capozziello:2022vyd}; see also \cite{Cai:2015emx} for a review.\footnote
	{We remark that problems associated with strong coupling were reported in some of these models \cite{Izumi:2012qj, Ong:2013qja,Jimenez:2019woj,BeltranJimenez:2019tme,Jimenez:2021hai}.}

\begin{figure}[H]
		\centering
		\includegraphics[width=0.5\linewidth]{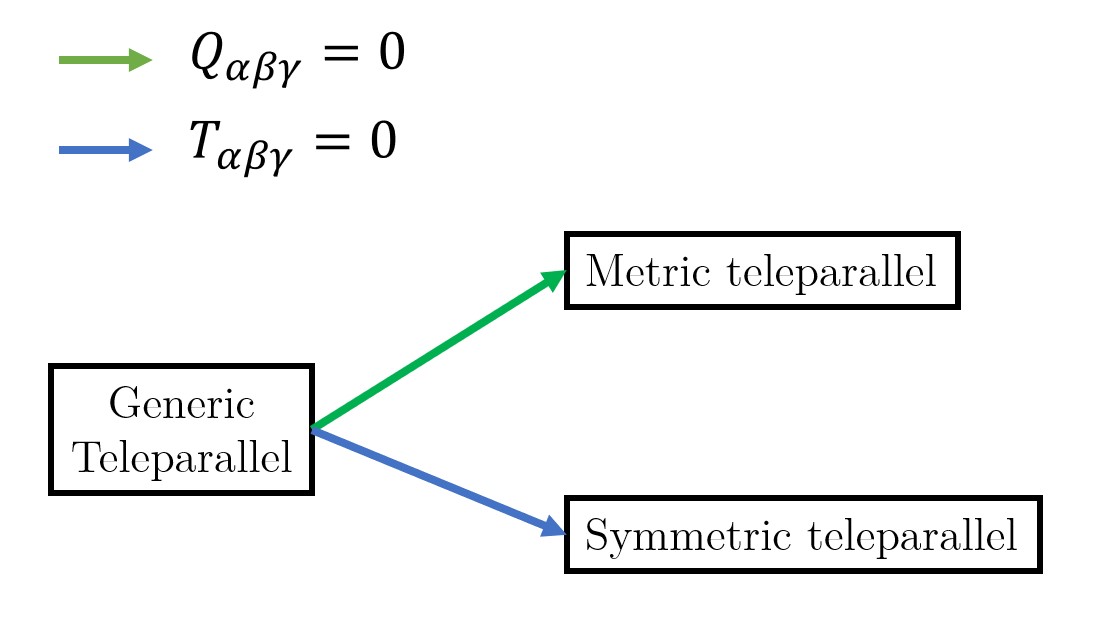}
		\caption{Diagram of relations between different formulations of gravity without curvature. Starting from the most general class of teleparallel theories of gravity on the left, one can obtain specific limits by imposing that torsion or non-metricity vanish.}
		\label{schema_theories_without_curvature}
	\end{figure}

	A summary of the relation between the different theories of gravity without curvature is given in \fig \ref{schema_theories_without_curvature}.
	
	\subsection{Selection rules}
	\label{ssec:selectionrules}
	
	In the preceding section, we have already outlined schematically the class of models that we shall investigate. In order to proceed systematically, we will now review the criteria developed in \cite{Karananas:2021zkl} for constructing an action of matter coupled to gravity. Only torsion was considered in \cite{Karananas:2021zkl} but the conditions equally well apply to a metric-affine formulation in which both torsion and non-metricity are present. In addition to (implicit) requirements of Lorentz invariance and locality, the criteria of \cite{Karananas:2021zkl} demand the following:
	\begin{enumerate}
		\item The purely gravitational part of the action should only feature operators of mass dimension not greater than $2$.
		\item The matter Lagrangian should be renormalizable in the flat space limit, \ie for $g_{\mu\nu}= \eta_{\mu\nu}$ and $\Gamma_{~~\mu \lambda}^\alpha = 0$.
		\item The interaction of gravity and matter should only happen via operators of mass dimension not greater than $4$. 
	\end{enumerate}
	Subsequently, we shall discuss their motivation and implications.
	
	Since torsion and non-metricity have mass dimension $1$ and curvature has mass dimension $2$, criterion 1.) implies that terms at most quadratic in torsion/non-metricity and linear in curvature can be included. Following the arguments of the preceding section, one can equivalently say that this condition arises from the decomposition \eqref{curvatureSplit} of curvature. Namely, it amounts to including contributions analogous to those already contained in curvature but with arbitrary coefficients. The purpose of criterion 1.) is to ensure equivalence with metric GR in the absence of matter. Correspondingly, it excludes terms that are quadratic or higher in curvature since they generically lead to new propagating degrees of freedom. What is more, some of these additional particles also cause inconsistencies, in particular since they correspond to ghosts, \ie have a kinetic term with a wrong sign.
	Note that this is already the case in metric gravity \cite{Stelle:1977ry}, and numerous additional problematic terms arise in the presence of torsion \cite{Neville:1978bk,Neville:1979rb,Sezgin:1979zf,Hayashi:1979wj, Hayashi:1980qp}.
	We must mention, however, that certain combinations of curvature-squared contribution only lead to new propagating degrees of freedom that are healthy; see \cite{Sezgin:1981xs,Kuhfuss:1986rb,Yo:1999ex,Yo:2001sy,Nair:2008yh,Nikiforova:2009qr,Karananas:2014pxa,Karananas:2016ltn,Obukhov:2017pxa,Blagojevic:2017ssv,Blagojevic:2018dpz,Lin:2018awc,Jimenez:2019qjc,Lin:2019ugq} for studies in the presence of torsion and \cite{BeltranJimenez:2019acz,Aoki:2019rvi,Percacci:2019hxn,BeltranJimenez:2020sqf,Marzo:2021esg,Marzo:2021iok,Baldazzi:2021kaf} for extensions to non-metricity. Moreover, it is possible to construct theories with terms that are quadratic in curvature that do not feature at all any additional propagating degrees of freedom \cite{Karananas:2021zkl}. Therefore, criterion 1.) is sufficient but not strictly necessary for ensuring that the gravitational theory is equivalent to metric GR in pure gravity. 
	
	Criterion 2.) implies that the matter sector only contains operators of mass dimension not greater than $4$. This assumption is crucial for the predictiveness of our setup. Without it, one could have added from the beginning generic higher-dimensional operators to our model and the specific higher-dimensional operators that arise due to torsion and non-metricity would be meaningless. Needless to say, the validity of this approach, in which the matter Lagrangian solely features those non-renormalizable operators that result from torsion and non-metricity, remains to be checked. At least in principle, this can be done by systematically exploring the predictions that result from the specific higher-dimensional interactions and then comparing them with observations and experiments. In the present paper, we lay the groundwork for such studies by explicitly deriving the set of predicted operators in the scalar sector.
	
	Criterion 3.) can be regarded as an attempt to define the notion of non-minimal coupling independently of the formulation of GR. In metric gravity, there is a unique operator for coupling a $Z_2$-symmetric scalar field $h$ non-minimally to GR, namely $h^2 \mathring{R}$ (see \eq \eqref{actionMetricGravityMatterSymbolical}). Since it has mass dimension 4, criterion 3.) aims at generalizing the notion of non-minimal coupling by selecting all terms that are on the same footing as the non-minimal coupling in metric GR. However, criterion 3.) is not crucial for our approach. It can be relaxed, as long as one makes sure that the coupling of matter and gravity does not lead to any additional propagating degrees of freedom. Correspondingly, we shall keep our discussion general and not impose criterion 3.) in some parts of the present work. It will only be implemented from section \ref{subsec:interaction} on.
	
	\section{Scalar field coupled to metric-affine gravity}
	\label{sec:Gen-Action}

	\subsection{The theory}
	Next, we shall consider a scalar field $h$ coupled to gravity and write down the most general action obeying selection rules 1.) and 2.). We will rely on a decomposition of torsion and non-metricity into vector- and tensor-parts, as shown in \eqs \eqref{torsionTrace} to \eqref{irrep_metricity}. This method, which was introduced in \cite{Obukhov:1997zd}, makes it significantly easier to solve the equations of motion. We get the action 
	\begin{subequations} \label{general_action_component}
		\begin{align}
			S=&\int \mathrm{d}^{4} x \sqrt{-g}\Big[\frac{1}{2} \Omega^2(h) \mathring{R} -\frac{1}{2} \tilde{K}(h) g^{\alpha \beta} \partial_{\alpha} h \partial_{\beta} h-V(h)\\
			& +A_1(h) \mathring{\nabla}_\alpha \hat{T}^\alpha +A_2(h) \mathring{\nabla}_\alpha T^\alpha+A_3(h) \mathring{\nabla}_\alpha \hat{Q}^\alpha +A_4(h) \mathring{\nabla}_\alpha Q^\alpha\\
			&+B_{1}(h) Q_{\alpha} Q^{\alpha}+B_{2}(h) \hat{Q}_{\alpha} \hat{Q}^{\alpha}+B_{3}(h) Q_{\alpha} \hat{Q}^{\alpha}+B_{4}(h) q_{\alpha \beta \gamma} q^{\alpha \beta \gamma}+B_{5}(h)q_{\alpha \beta \gamma} q^{\beta \alpha \gamma}\\
			&+C_{1}(h) T_{\alpha} T^{\alpha}+C_{2}(h) \hat{T}_{\alpha} \hat{T}^{\alpha}+C_{3}(h) T_{\alpha} \hat{T}^{\alpha} +C_{4}(h)t_{\alpha \beta \gamma}t^{\alpha \beta \gamma}\\
			&+D_1(h)\epsilon_{\alpha \beta \gamma \delta}t^{\alpha \beta \lambda}t^{\gamma \delta}_{~~\lambda}+D_2(h)\epsilon_{\alpha \beta \gamma \delta}q^{\alpha \beta \lambda}q^{\gamma \delta}_{~~\lambda}+D_{3}(h)\epsilon_{\alpha \beta \gamma \delta}q^{\alpha \beta \lambda}t^{\gamma \delta}_{~~\lambda}\\
			&+E_1(h) T_{\alpha}Q^\alpha+E_2(h)\hat{T}_{\alpha}Q^\alpha+E_3(h) T_{\alpha}\hat{Q}^\alpha+E_4(h)\hat{T}_\alpha \hat{Q}^\alpha+E_5(h)t^{\alpha \beta \gamma}q_{\beta \alpha \gamma}\Big] .
		\end{align}
	\end{subequations}
	Since at this point we have not yet enforced selection rule 3.) of section \ref{ssec:selectionrules}, $\Omega^2(h)$, $\tilde{K}(h)$, $A_i(h)$, $B_i(h)$, $C_i(h)$, $D_i(h)$ and $E_i(h)$ in \eq \eqref{general_action_component} represent arbitrary functions of $h$. Besides, some of the possible non-vanishing terms have not been included in the action, such as \eg $t_{\alpha \beta \gamma}t^{\beta \alpha \gamma} $. The reason is that they are linearly dependent on terms that are already present. For more details, we refer the reader to appendix \ref{app:independence_terms}, where the independence of terms is discussed. 
	Notice that for generic choices of coefficient functions, the action \eq \eqref{general_action_component} is not invariant under the projective transformation shown in \eq \eqref{projective}. Thus, the connection can be uniquely determined by its equations of motion.
	
	Let us briefly comment on related works in metric-affine gravity. A general action that features all independent invariants composed of torsion and non-metricity was already introduced in \cite{Pagani:2015ema}, where torsion and non-metricity were not split into pure vector- and tensor-parts. The matter sector was not made explicit in \cite{Pagani:2015ema}, and so the functions $\Omega^2(h)$, $\tilde{K}(h)$ and $A_i(h)$ were not present.  The action proposed in \cite{Pagani:2015ema} was further studied in \cite{Iosifidis:2021bad} and solutions for torsion and non-metricity were derived in terms of energy-momentum- and hypermomentum-tensors. Unlike in the present work, the paper \cite{Iosifidis:2021bad} employed a method for finding solutions that does not require the separation of pure tensor parts \cite{Iosifidis:2021ili, Iosifidis:2021kwd}.
	
	\subsection{Derivation of equivalent metric theory}
	We can now vary the action given in \eq \eqref{general_action_component} with respect to the six tensors  $T^{\alpha}$, $\hat{T}^{\alpha}$, $t^{\alpha \beta \gamma}$, $Q^\gamma$, $\hat{Q}^\gamma$ and $q^{\alpha \beta \gamma}$, as discussed in section \ref{ssec:decomposition}. We obtain the following equations of motion: 
	\begin{equation}
		\begin{split}
			&2 C_2(h) \hat{T}^\alpha+C_3(h)T^\alpha+E_2(h) Q^\alpha+E_4(h) \hat{Q}^\alpha= A_1'(h) \partial^\alpha h \;, \\
			&2 C_1(h) T^\alpha+C_3(h)\hat{T}^\alpha+E_1(h)Q^\alpha+E_3(h)\hat{Q}^\alpha= A_2'(h) \partial^\alpha h \;, \\
			&2 B_2(h)\hat{Q}^\alpha+B_3(h)Q^\alpha+E_3(h) T^\alpha+E_4(h)\hat{T}^\alpha= A_3'(h) \partial^\alpha h \;, \\
			&2B_1(h) Q^\alpha+B_3(h) \hat{Q}^\alpha+E_1(h) T^\alpha+E_2(h) \hat{T}^\alpha= A_4'(h)\partial^\alpha h \;,
			\\&2 B_4(h) q_{\alpha \beta \gamma}+ 2B_5(h)q_{(\beta \alpha) \gamma} +2D_2(h) \epsilon_{\alpha \lambda \delta (\beta}q^{\lambda \delta}_{~~\gamma )}+D_3(h) \epsilon_{\alpha \lambda \delta (\beta} t^{\lambda\delta}_{~~\gamma)}-E_5(h) t_{(\beta \gamma) \alpha}=0 \;,\\
			&2C_4(h) t_{\alpha \beta \gamma}+2D_1(h)		\epsilon_{\alpha \lambda \delta [\beta}t^{\lambda \delta}_{~~\gamma ]}+D_3(h)\epsilon_{\alpha \lambda \delta [\beta}q^{\lambda \delta}_{~~\gamma]}+E_5(h)q_{[\beta \gamma]\alpha}=0 \;,
		\end{split}
	\end{equation}
	where prime denotes derivative with respect to $h$. Solutions can be found explicitly as the equations of motion are algebraic. We first notice that there are no source terms for the pure tensor parts, hence they simply vanish.\footnote
	{This is related to the fact that there is no Lorentz-invariant derivative of a pure tensor part with mass dimension not greater than $2$. If we were to relax the first selection rule imposed in section \ref{ssec:selectionrules}, then it would be possible to write terms like $F(h)T_\alpha \hat{T}_\beta \mathring{\nabla}_\gamma t^{\alpha \beta \gamma}$ which could act as source terms.}
	On the contrary, the vector parts $T^\alpha$, $\hat{T}^\alpha$, $Q^\alpha$ and $\hat{Q}^\alpha$ do not vanish because of the presence of source terms $A_i(h)$. We obtain the solutions:\footnote{For simplicity, we removed the explicit dependence on the scalar field $h$.}: 
	\begin{equation}
		Q^\alpha=\frac{V}{Z}\partial^\alpha h \;, \qquad \hat{Q}^\alpha=\frac{W}{Z}\partial^\alpha h \;, \qquad 	T^\alpha=\frac{X}{Z}\partial^\alpha h \;, \qquad \hat{T}^\alpha=\frac{Y}{Z}\partial^\alpha h \;, \qquad t_{\alpha \beta \gamma}=q_{\alpha \beta \gamma}=0 \;.
		\label{solution_short}
	\end{equation}
	The common denominator reads
	\begin{equation}
		\begin{split}
			Z&=B_3^2 (4 C_1 C_2 - C_3^2) + 4 B_2 C_2 E_1^2 - 4 B_2 C_3 E_1 E_2 + 4 B_2 C_1 E_2^2 - 
			E_2^2 E_3^2 + 2 E_1 E_2 E_3 E_4 \\&- E_1^2 E_4^2 +
			B_3 (-4 C_2 E_1 E_3 + 2 C_3 E_2 E_3 + 2 C_3 E_1 E_4 - 4 C_1 E_2 E_4) + 
			4 B_1 (B_2 (-4 C_1 C_2 + C_3^2)\\& + C_2 E_3^2 - C_3 E_3 E_4 + C_1 E_4^2) \;,
		\end{split}
	\end{equation}
	and the numerators are:
	\begin{equation}
		\begin{split}
			V&= 4 A'_2 B_2 C_2 E_1 - 2 A'_1 B_2 C_3 E_1 + 4 A'_1 B_2 C_1 E_2 - 2 A'_2 B_2 C_3 E_2 - 
			2 A'_2 B_3 C_2 E_3 + A'_1 B_3 C_3 E_3 \\&- A'_1 E_2 E_3^2 - 2 A'_1 B_3 C_1 E_4 + 
			A'_2 B_3 C_3 E_4 + A'_1 E_1 E_3 E_4 + A'_2 E_2 E_3 E_4 - A'_2 E_1 E_4^2\\& + 
			A'_3 (4 B_3 C_1 C_2 - B_3 C_3^2 - 2 C_2 E_1 E_3  + C_3 E_2 E_3 + C_3 E_1 E_4 - 
			2 C_1 E_2 E_4) \\&+ 
			2 A'_4 (B_2 (-4 C_1 C_2 + C_3^2) + C_2 E_3^2 - C_3 E_3 E_4 + C_1 E_4^2) \;,
		\end{split}
	\end{equation}
	\begin{equation}
		\begin{split}
			W&=-2 A'_2 B_3 C_2 E_1 + A'_1 B_3 C_3 E_1 - 2 A'_1 B_3 C_1 E_2 + A'_2 B_3 C_3 E_2 + 
			2 A'_3 (B_1 (-4 C_1 C_2 + C_3^2) + C_2 E_1^2\\& - C_3 E_1 E_2 + C_1 E_2^2) + 
			4 A'_2 B_1 C_2 E_3 - 2 A'_1 B_1 C_3 E_3 + A'_1 E_1 E_2 E_3 - A'_2 E_2^2 E_3 + 
			4 A'_1 B_1 C_1 E_4\\& - 2 A'_2 B_1 C_3 E_4 - A'_1 E_1^2 E_4 + A'_2 E_1 E_2 E_4 + 
			A'_4 (4 B_3 C_1 C_2 - B_3 C_3^2 - 2 C_2 E_1 E_3 + C_3 E_2 E_3 \\&+ C_3 E_1 E_4 - 
			2 C_1 E_2 E_4)  \;,
		\end{split}
	\end{equation}
	\begin{equation}
		\begin{split}
			X&=4 A'_4 B_2 C_2 E_1 - 2 A'_3 B_3 C_2 E_1 - 2 A'_4 B_2 C_3 E_2 + A'_3 B_3 C_3 E_2 + 
			4 A'_3 B_1 C_2 E_3 - 2 A'_4 B_3 C_2 E_3 \\&- A'_3 E_2^2 E_3 - 2 A'_3 B_1 C_3 E_4 + 
			A'_4 B_3 C_3 E_4 + A'_3 E_1 E_2 E_4 + A'_4 E_2 E_3 E_4 - A'_4 E_1 E_4^2 \\&+ 
			A'_1 (4 B_1 B_2 C_3 - B_3^2 C_3 - 2 B_2 E_1 E_2 + B_3 E_2 E_3 + B_3 E_1 E_4 - 
			2 B_1 E_3 E_4) \\&+ 
			2 A'_2 (B_3^2 C_2 + B_2 E_2^2 - B_3 E_2 E_4 + B_1 (-4 B_2 C_2 + E_4^2))  \;,
		\end{split}
	\end{equation}
	\begin{equation}
		\begin{split}
			Y&= -2 A'_4 B_2 C_3 E_1 + A'_3 B_3 C_3 E_1 + 4 A'_4 B_2 C_1 E_2 - 2 A'_3 B_3 C_1 E_2 - 
			2 A'_3 B_1 C_3 E_3 + A'_4 B_3 C_3 E_3 \\&+ A'_3 E_1 E_2 E_3 - A'_4 E_2 E_3^2 + 
			2 A'_1 (B_3^2 C_1 + B_2 E_1^2 - B_3 E_1 E_3 + B_1 (-4 B_2 C_1 + E_3^2)) + 
			4 A'_3 B_1 C_1 E_4 \\&- 2 A'_4 B_3 C_1 E_4 - A'_3 E_1^2 E_4 + A'_4 E_1 E_3 E_4 + 
			A'_2 (4 B_1 B_2 C_3 - B_3^2 C_3 - 2 B_2 E_1 E_2 + B_3 E_2 E_3 \\&+ B_3 E_1 E_4 - 
			2 B_1 E_3 E_4) \;.
		\end{split}
	\end{equation}
	The expressions for the numerators and denominator are quite long but the overall form of the solution for torsion and non-metricity is simple: they are proportional to $\partial^\alpha h$, as shown in \eq\eqref{solution_short}.
	
	The fact that the pure tensor parts $t^{\alpha\beta \gamma}$ and $q^{\alpha \beta \gamma}$ vanish dynamically has a remarkable consequence. As is evident from \eq \eqref{general_action_component}, the similarity between terms containing torsion and terms containing non-metricity is only broken because of the pure tensor parts and their different symmetry properties. Once $t^{\alpha\beta \gamma}$ and $q^{\alpha \beta \gamma}$ are absent, however, an exact correspondence emerges between a theory that only features torsion and a model that solely contains non-metricity. Thus, our criteria for the construction of an action of gravity coupled to matter lead to a full equivalence of the Einstein-Cartan and Weyl formulations. We will make this point explicit in section \ref{sec:knownlimits}.

	Summarizing what we did so far, we started from the most general action \eqref{general_action_component} according to the criteria presented in section \ref{ssec:selectionrules}. Torsion and non-metricity are included, therefore we can write new terms that are absent in the metric formulation of GR. We then solved for both torsion and non-metricity and found that they are proportional to the derivative of the scalar field. As next step, we can plug these solutions back into the action \eq \eqref{general_action_component}. Then the new terms will give contributions to the kinetic term of the scalar field, \ie we can map the effect of torsion and non-metricity to a modification of the kinetic term. We get 
	\begin{equation} 
		\begin{aligned}
			S=&\int \mathrm{d}^{4} x \sqrt{-g}\left[\frac{1}{2} \Omega^2(h) \mathring{R}-\frac{1}{2} \hat{K}(h) g^{\alpha \beta} \partial_{\alpha} h \partial_{\beta} h-V(h)\right] \;,
		\end{aligned}
		\label{action_mid}
	\end{equation}
	where the modified kinetic term reads: 
	\begin{equation}
		\begin{split}
			\hat{K} = \tilde{K}-2\frac{1}{Z^2}[&B_1 V^2+B_2W^2+B_3VW+C_1 X^2+C_2Y^2+C_3X Y+E_1 VX +E_2VY\\ &+E_3WX+E_4WY -Z(A'_1 Y +A'_2 X +A'_3 W +A'_4V)] \;.
			\label{modified_kinetic_term}
		\end{split}
	\end{equation}
	Since our result only features the torsion- and non-metricity-free curvature $\mathring{R}$, which is fully determined in terms of the metric $g_{\mu\nu}$, we can call \eq \eqref{action_mid} the \textit{equivalent metric theory}: We are back to a situation where the metric $g_{\mu \nu}$ is the only degree of freedom in the gravity sector.
	
	At this point we  are still in the Jordan frame where the Ricci scalar $\mathring{R}$ is multiplied by $\Omega^2(h)$, meaning that the scalar field is non-minimally coupled to gravity. One may perform a conformal transformation in order to go to the Einstein frame, where the coupling to gravity is minimal \cite{Carroll:2004}:
	\begin{equation}
		\begin{split}
			&g_{\alpha \beta} \rightarrow \Omega^{-2} g_{\alpha \beta} \;, \\
			& g^{\alpha \beta} \rightarrow \Omega^{2} g^{\alpha \beta} \;,\\
			& \sqrt{-g} \rightarrow \Omega^{-4} \sqrt{-g} \;, \\
			&g^{\alpha \beta} \mathring{R}_{\alpha \beta} \rightarrow \Omega^{2}[g^{\alpha \beta} \mathring{R}_{\alpha \beta}+6g^{\alpha \beta}(\mathring{\nabla}_\alpha \mathring{\nabla}_\beta \ln(\Omega)-\mathring{\nabla}_\alpha \ln(\Omega)\mathring{\nabla}_\beta \ln(\Omega))] \;.\\
		\end{split}
		\label{158}
	\end{equation}
	Notice that the scalar curvature $\mathring{R}$  transforms inhomogeneously due to the dependence of the Levi-Civita connection $\mathring{\Gamma}^\gamma_{~\alpha \beta}$ on the metric $g_{\mu\nu}$. This inhomogeneous contribution leads to another modification of the kinetic term of the scalar field $h$.\footnote
	{Notice that under the conformal transformation the derivative of the metric changes, $ \nabla_\mu g_{\alpha \beta} \rightarrow \nabla_\mu(\Omega^{-2}g_{\alpha \beta})$, which implies that the non-metricity tensor $Q_{\alpha \beta \gamma}$ transforms inhomogeneously. For our discussion this is inessential since $Q_{\alpha \beta \gamma}$ does not appear any more in the action \eqref{action_mid}.}
	After the conformal transformation, the action takes the form: 
	\begin{equation}
		\begin{aligned}
			S=&\int \mathrm{d}^{4} x \sqrt{-g}\left[\frac{1}{2} \mathring{R}-\frac{1}{2} K(h) g^{\alpha \beta} \partial_{\alpha} h \partial_{\beta} h-\frac{V(h)}{\Omega^4}\right] \;.
			\label{action_equivalent_metric}
		\end{aligned}
	\end{equation}
	The kinetic function in the Einstein frame is:  
	\begin{equation}
		\begin{split}
			K&=\frac{\hat{K}}{\Omega^2}+ \frac{6(\Omega'(h))^2}{\Omega^2}\\
			&= \frac{1}{\Omega^2}[\tilde{K}-2\frac{1}{Z^2}\{B_1 V^2+B_2W^2+B_3VW+C_1 X^2+C_2Y^2+C_3X Y+E_1 VX +E_2VY\\ & \qquad +E_3WX+E_4WY -Z(A'_1 Y +A'_2 X +A'_3 W +A'_4V)\}+6(\Omega'(h))^2] \;,
		\end{split}
		\label{230}
	\end{equation}
	where $\Omega'(h)$ denotes the derivative of the function with respect to $h$. It is evident from \eq \eqref{action_equivalent_metric} that the effect of non-minimal coupling to $\mathring{R}$ is mapped to the kinetic term of the scalar as well as to a modification of the potential of the scalar field.

	\subsection{Interaction between matter and gravity sectors}\label{subsec:interaction}
	Finally, we will impose criterion 3.) from section \ref{ssec:selectionrules}. In this way, we reduce the functional freedom present in \eq \eqref{general_action_component} to a finite number of coupling constants. Moreover, we shall assume that the scalar field $h$ obeys a $\mathbb Z^2$ symmetry $h\rightarrow -h$. This condition is motivated by the fact that the Higgs field of the Standard Model exhibits the same property in unitary gauge. Apart from the $Z_2$-symmetry, however, the scalar field in the present paper is generic and does not need to represent the Higgs boson. We get 
	\begin{equation}
		\begin{split}
			&\tilde{K}(h)=k_0 \;, \qquad  \Omega^2(h)=f_0+\xi h^2 \;, \qquad D_i(h)=d_{i0}+d_{i1}h^2\,, \qquad i=1,2,3 \;, 
			\\&  A_j(h)=a_{j1}h^2 \;, \qquad 
			C_j(h)=c_{j0}+c_{j1}h^2 \;,  \qquad j=1,2,3,4 \;,\\
			& B_k(h)=b_{k0}+b_{k1}h^2 \;, \qquad E_k(h)=e_{k0}+e_{k1}h^2 \;, \qquad  k=1,2,3,4,5 \;.
		\end{split}
		\label{selec_rule_3}
	\end{equation}
	Without loss of generality, one can set $f_0=k_0=1$ by a redefinition of the scalar field and rescalings of the other parameters of the theory (including those contained in $V(h)$).\footnote{As becomes apparent in \eq \eqref{kinetic_after_conformal}, the effects of all parameters except for $f_0$ and $k_0$ are suppressed at small energies. The choice $f_0=k_0=1$ ensures that in the limit of small field values, the scalar field $h$ and gravitational perturbations $h_{\mu\nu}$, defined by $g_{\mu\nu} = \eta_{\mu\nu}+ h_{\mu\nu}/M_P$, are already canonically normalized.} At this point we have $39$ independent couplings in the action: $1$ for $\xi$ and $38$ coming from the terms in the functions $A_i$, $B_i$, $C_i$, $D_i$ and $E_i$.
	
	The kinetic term \eqref{modified_kinetic_term}, \ie before the conformal transformation, becomes
\begin{equation}
	\hat{K}(h)= 1 +\frac{h^2}{ \sum_{m=0}^4 O_m h^{2m}}\sum_{n=0}^3 P_n h^{2n} \;,
	\label{kinetic_before_conformal}
\end{equation}
where $O_m$ and $P_n$ are polynomials of the constants defined in \eq \eqref{selec_rule_3}. Their explicit expressions are lengthy  (up to a few pages) and will not be displayed.
	After the conformal transformation, the kinetic function in the Einstein frame action \eqref{action_equivalent_metric} is: 
\begin{equation}
		K(h)=\frac{1}{(1+\xi h^2)}\left[1+\frac{h^2}{ \sum_{m=0}^4 O_m h^{2m}}\sum_{n=0}^3 P_n h^{2n} + \frac{6\xi^2h^2}{(1+\xi h^2)} \right] \;.
		\label{kinetic_after_conformal}
	\end{equation}
Inspecting the second summand in \eq\eqref{kinetic_after_conformal},  we see that there are $4$ independent polynomials in the numerator and $5$ in the denominator. Moreover, we have to take into account the parameter $\xi$. Finally, we need to effectively deduce one coupling constant since we can rescale numerator and denominator by a common factor. In total, this leads to $4+5+1-1=9$ independent constants, whereas there were $39$ previously. This shows that in the case of a single scalar field, torsion and non-metricity effects only depend on a subset of combinations of the initial constants and that there is redundancy.
	
	\subsection{Known limits as special cases of the general action}
	\label{sec:knownlimits}
	Let us show how the action proposed in \eq\eqref{general_action_component} reduces to different formulations of gravity. First we will prove how we can obtain Einstein-Cartan gravity (where torsion is present but non-metricity vanishes) by comparing explicitly expressions with \cite{Karananas:2021zkl}. Then we will discuss its similarities with Weyl formulation of gravity (where instead torsion vanishes but non-metricity is present). Finally we will compare it to a mixed theory proposed in \cite{Rasanen:2018ihz}.
	
	\paragraph{Einstein-Cartan gravity}
	We can obtain Einstein-Cartan gravity from the metric-affine formulation employed in the present paper by setting to zero all coefficients of terms that involve non-metricity: 
	\begin{equation}	\label{parametersEC2}	
		A_3=A_4=B_i=E_i=D_2=D_3=0 \;.	
	\end{equation}
	Then the kinetic term \eqref{230} becomes: 
	\begin{equation}\label{kin_EC_before_correspondence}
		K_{EC}=\frac{1}{\Omega^2}\left[\tilde{K}+\frac{2C_1(A'_1)^2+2C_2(A'_2)^2-2C_3A'_1A'_2}{4C_1C_2-C_3^2}+6(\Omega')^2\right] \;.
	\end{equation}
	In this way, we can reproduce the result of \cite{Karananas:2021zkl}. In turn, \cite{Karananas:2021zkl} encompasses numerous previous studies as special cases such as \cite{Perez:2005pm, Freidel:2005sn,Alexandrov:2008iy,Taveras:2008yf, Torres-Gomez:2008hac, Calcagni:2009xz,Mercuri:2009zi, Diakonov:2011fs, Magueijo:2012ug,Langvik:2020nrs, Shaposhnikov:2020frq}. The correspondence between \eq \eqref{general_action_component} and the action in \cite{Karananas:2021zkl} is given by : 
	\begin{equation} \label{parametersEC}
		\begin{split}
			& \tilde{K}=1 \;, \qquad V=U \;, \qquad A_1=-Z^a \;, \qquad A_2=-Z^v \;, \qquad C_1=\frac{1}{2}G_{vv} \;, \\ &C_2=\frac{1}{2}G_{aa} \;, \qquad C_3=G_{va} \;, \qquad C_4=\frac{1}{2} G_{\tau \tau} \;, \qquad D_1=2\tilde{G}_{\tau \tau} \;.
		\end{split}
	\end{equation}
	Plugging this in \eq \eqref{kin_EC_before_correspondence}, we obtain 
	\begin{equation}
		K_{EC}=\frac{1}{\Omega^2}\left[1+\frac{G_{vv}(Z^{a \prime})^2+G_{aa}(Z^{v \prime})^2-2G_{va}Z^{v \prime} Z^{a \prime}}{G_{vv} G_{aa}-G_{va}^2}+6(\Omega')^2\right] \;,
	\end{equation}
	matching what is found in \cite{Karananas:2021zkl}. We can expand the functions like in \eq\eqref{selec_rule_3} by imposing selection rule 3.) and the final result for the kinetic term in the Einstein frame is:
	\begin{equation} \label{KEC}
		K_{EC}(h)=\frac{1}{(1+\xi h^2)}\left[1+\frac{8 h^2}{\sum_{m=0}^2 \tilde{O}_m h^{2m}}\sum_{n=0}^1 \tilde{H}_n h^{2n} + \frac{6\xi^2h^2}{(1+\xi h^2)}\right] \;,
	\end{equation}
	where $\tilde{H}_n$ and $\tilde{O}_m$  are functions of the coefficient given by: 
	\begin{equation} 
		\begin{split}
			&\tilde{H}_0=a_{11}^2 c_{10} + a_{21}^2 c_{20}-a_{11} a_{21} c_{30} \;, \qquad \tilde{H}_1=a_{11}^2 c_{11} + a_{21}^2 c_{21}-a_{11} a_{21} c_{31}\;, \\
			&\tilde{O}_0=4 c_{10} c_{20}-c_{30}^2 \;, \qquad \tilde{O}_1=4 c_{11} c_{20}+4 c_{10} c_{21}-2 c_{30} c_{31}\;, \qquad \tilde{O}_2= 4 c_{11} c_{21}-c_{31}^2\;.
		\end{split}
	\end{equation}
	Eq.\@\xspace \eqref{KEC} shows that there are $2$ independent polynomials in the numerator and $3$ in the denominator. As before, we have the additional parameter $\xi$ of the non-minimal coupling to curvature $\mathring{R}$ but it is effectively canceled since we can rescale numerator and denominator by a common factor. In total, we obtain $2 + 3 + 1 -1 = 5$ independent parameters. We can contrast this with $9$ independent polynomials in the general case shown in \eq \eqref{kinetic_after_conformal}. Einstein-Cartan gravity is indeed a very specific limit of the general metric-affine theory. 
	
	\paragraph{Comparison of Einstein-Cartan and Weyl gravity}
	Weyl gravity is the counterpart of Einstein-Cartan gravity: torsion is assumed to vanish a priori but non-metricity is present. This leads to the following simplifications in action \eqref{general_action_component}: 
	\begin{equation}
		A_1=A_2=C_i=D_1=D_3 =E_i=0 \;.
	\end{equation}
	Plugging these constraints into the modified kinetic term \eq \eqref{230}, we find 
	\begin{equation}
		K_{Weyl}=\frac{1}{\Omega^2}\left[\tilde{K}+\frac{2B_1(A'_3)^2+2B_2(A'_4)^2-2B_3A'_3A'_4}{4B_1B_2-B_3^2}+6(\Omega')^2\right] \;.
		\label{kin_weyl}
	\end{equation}
	This result is identical to the kinetic term \eqref{kin_EC_before_correspondence} in the Einstein-Cartan case, after the identifications
	\begin{equation} \label{identification_weyl_EC}
		C_i \longleftrightarrow B_i , \qquad A_1 \longleftrightarrow A_3, \qquad A_2 \longleftrightarrow A_4 \;.
	\end{equation} 
	As previously discussed, the Einstein-Cartan and Weyl formulations are equivalent for the choice \eqref{general_action_component} of action.
	
	\paragraph{Mixed theory with torsion and non-metricity}
	Finally, we demonstrate that the action of a mixed theory, as given in \cite{Rasanen:2018ihz}, also represents a special case of our metric-affine model. The action is \cite{Rasanen:2018ihz}: 
	\begin{equation} \label{actionSyksy}
		\begin{aligned}
			S_{\text{mixed}}= &\int \mathrm{d}^{4} x \sqrt{-g}\left[\frac{1}{2} F(h) R-\frac{1}{2} \tilde{K}(h) g^{\alpha \beta} \nabla_{\alpha} h \nabla_{\beta} h-V(h)\right. \\
			&-\tilde{A}_{1}(h) \nabla_{\alpha} h \hat{Q}^{\alpha}-\tilde{A}_{2}(h) \nabla_{\alpha} h Q^{\alpha} \\
			&+\tilde{B}_{1}(h) Q_{\gamma \alpha \beta} Q^{\gamma \alpha \beta}+\tilde{B}_{2}(h) Q_{\gamma \alpha \beta} Q^{\beta \gamma \alpha}+\tilde{B}_{3}(h) \hat{Q}_{\alpha} \hat{Q}^{\alpha}+\tilde{B}_{4}(h) Q_{\alpha} Q^{\alpha}+\tilde{B}_{5}(h) Q_{\alpha} \hat{Q}^{\alpha} \\
			&\left.+\tilde{C}(h) \epsilon^{\alpha \beta \gamma \delta} g^{\epsilon \eta} Q_{\alpha \gamma \epsilon} Q_{\beta \delta \eta}\right] .
		\end{aligned}
	\end{equation}
	To be able to make the comparison, we need to decompose the scalar curvature $R$ as well as the terms $Q_{\gamma \alpha \beta} Q^{\gamma \alpha \beta}$ and $Q_{\gamma \alpha \beta} Q^{\beta \gamma \alpha}$ into contributions of vectors and pure tensors. Using \eqs \eqref{curvatureSplit}, \eqref{Qdecomposition1} and \eqref{Qdecomposition2}, we obtain the correspondence: 
	\begin{equation}
		\begin{split}
			& \Omega^2=F\;, \qquad  A_1'=0\;, \qquad A_2'=F'\;, \qquad A_3'=\tilde{A}_1-\frac{F'}{2}\;, \qquad A_4'=\tilde{A}_2+\frac{F'}{2}\;,\\&   B_1=\frac{5}{18}\tilde{B}_1-\frac{1}{18}\tilde{B}_2+\tilde{B}_4-\frac{11}{144}F \;, \qquad B_2=\frac{4}{9}\tilde{B}_1+\frac{1}{9}\tilde{B}_2+\tilde{B}_3+\frac{1}{36}F\;,\\& B_3=-\frac{2}{9}\tilde{B}_1+\frac{4}{9}\tilde{B}_2+\tilde{B}_5+\frac{1}{9}F\;, \qquad B_4= \tilde{B}_1+\frac{1}{8}F\;, \qquad B_5=\tilde{B}_2-\frac{1}{4}F\;, \qquad C_1=-\frac{1}{3}F\;,  \\& C_2=\frac{F}{48}\;, \qquad C_3=0 \;, \qquad C_4=\frac{1}{4}F\;,\qquad D_1=0\;,\qquad D_2=-\tilde{C}\;,\qquad D_3=0\;, \qquad D_4=0\;, \\& E_1=-\frac{F}{3}\;, \qquad E_2=0\;, \qquad E_3=\frac{F}{3}\;, \qquad E_4=0\;, \qquad E_5=\frac{F}{2} \;.
		\end{split}
	\end{equation}
	Plugging this into the kinetic term \eqref{230} after the conformal transformation yields: 
	\begin{equation}
		\begin{split}
			K=\frac{\tilde{K}}{F} &+ \frac{1}{2 F M} \Big( F (\tilde{A_1}+4 \tilde{A_2})^2 + 8 \tilde{A_1}^2 (5 \tilde{B_1}-\tilde{B_2}+18 \tilde{B_4})\\
			&+16 \tilde{A_1} \tilde{A_2} (2 \tilde{B_1}-4 \tilde{B_2}-9 \tilde{B_5})+16 \tilde{A_2}^2 (4 \tilde{B_1}+\tilde{B_2}+9 \tilde{B_3})\Big) \;,
		\end{split}
	\end{equation}
	where we defined (as in \cite{Rasanen:2018ihz}) 
	\begin{equation}
		\begin{split}
			M&= 16 \tilde{B_1}^2 -8 \tilde{B_2}^2 -36 \tilde{B_5}^2 +4 \tilde{B_1} (2\tilde{B_2}+10 \tilde{B_3}+16 \tilde{B_4}+4 \tilde{B_5}) +144 \tilde{B_3} \tilde{B_4}\\
			&+ \tilde{B_2} (-8\tilde{B_3}+16 \tilde{B_4}-32 \tilde{B_5}) +F(4 \tilde{B_1}+\tilde{B_2}+\tilde{B_3}+16 \tilde{B_4}+4 \tilde{B_5}) \;.
		\end{split}
	\end{equation}
	This matches the result obtained in \cite{Rasanen:2018ihz}.\footnote
	{The kinetic function $K(h)$ is displayed in \eq (29) of \cite{Rasanen:2018ihz}, where \eq (27) needs to be plugged in. As confirmed after correspondence with Syksy Räsänen, there is a minor typo in \cite{Rasanen:2018ihz}: The very first line of \eq (25) should read $$K \rightarrow \tilde{K}=K-3 F\left(\Sigma_{1}^{2}+\Sigma_{2}^{2}+\Sigma_{3}^{2}+4 \Sigma_{1} \Sigma_{2}+2 \Sigma_{2} \Sigma_{3}+4 \Sigma_{3} \Sigma_{1}\right) .$$ Accordingly, the second line of \eq (29) should be modified to: 
		$$ +F\left(-18 \Sigma_{1}^{2}\right)-8\left[A_{1}+\left(2 B_{2}+2 B_{3}+4 B_{5}\right) \omega^{\prime}\right]\left(2 \Sigma_{1}+\Sigma_{2}\right).$$}
	Finally imposing selection criterion 3.), we obtain 
	\begin{equation} \label{kin_modified_after_selec_3}
		K(h)=\frac{1}{(1+\tilde{\xi} h^2)}\left[1+\frac{h^2}{\sum_{m=0}^2 F_m h^{2m}}\sum_{n=0}^1 G_n h^{2n}\right] \;,
	\end{equation}
	where we set $F = 1 + \tilde{\xi} h^2$. The symbol $\tilde{\xi}$ is used instead of $\xi$ to indicate that $F$ couples to the full Ricci scalar $R$ and not only the Levi-Civita part $\mathring{R}$. We conclude that we have $2$ independent polynomials in the numerator and $3$ in the denominator. Also taking into account $\tilde{\xi}$ and the common rescaling of numerator and denominator, we arrive at $2+3+1-1=5$ independent parameters. Comparison with \eq \eqref{KEC} shows that the kinetic function of a real scalar field in the model \eqref{actionSyksy} has the same number of independent polynomials as in pure Einstein-Cartan or pure Weyl gravity.
	
	\paragraph{Summary} A summary of the different numbers of independent couplings is shown in table \ref{summary_indep_terms}. Let us explain what we mean by independent couplings and provide an explicit example for the metric-affine theory of gravity. After imposing selection criterion 3.), our initial action \eqref{general_action_component} features $39$ coupling constants. 
	However the pure tensor parts of torsion and non-metricity will vanish. This makes $7$ terms vanishing, so the number of independent couplings reduces to $39-7\times 2=25$. Finally, once we have solved for torsion and non-metricity, we obtain an expression for the modified kinetic term of the scalar field given in \eq \eqref{kinetic_after_conformal}. From there we can read off the number of independent polynomials: $1+4+5-1=9$. An analogous counting can be performed for the Einstein-Cartan and Weyl formulations as well as the model of \cite{Rasanen:2018ihz}.
	
	\begin{table}[h]
		\centering
		\begin{tabular}{|c|c|c|c|c|}
			\hline
			Theory of gravity &\parbox{0.2\linewidth}{Independent couplings in the initial action} & \parbox{0.2\linewidth}{Independent couplings after using $t^{\alpha \beta \gamma}=q^{\alpha \beta \gamma}=0$} & \parbox{0.2\linewidth}{Independent parameters  in the final kinetic term} \\ \hline
			Metric-affine    & $39$                                                                                   & $25$                                                                                                                                    & $9$                                                                                        \\ \hline
			Einstein-Cartan   & $13$                                                                                   & $9$                                                                                                                                        & $5$                                                                                          \\ \hline
			Weyl            & $15$                                                                                   & $9$                                                                                                                                        & $5$                                                                                          \\ \hline
			Mixed theory      & $15$                                                                                   & $9$                                                                                                                                       & $5$                                                                                         \\ \hline
			Metric gravity   & $1$                                                                                   & $1$                                                                                                                                       & $1$                                                                                          \\ \hline
		\end{tabular}
		\caption{Summary of the number of independent couplings for different formulations of GR.}
		\label{summary_indep_terms}
	\end{table}
	\section{Outlook: implications for Higgs inflation}
	\label{sec:HiggsInflation}
	In the following, we will discuss implications of our result, which is displayed in \eqs \eqref{action_equivalent_metric} and \eqref{230}, for Higgs inflation \cite{Bezrukov:2007ep}. To this end, we need to specify the potential $V(h)$. At large field values, which are relevant for inflation, we can neglect the electroweak vacuum expectation value of the Higgs field and approximate:
	\begin{equation}
		V(h) = \frac{\lambda}{4} h^4 \;,
	\end{equation}
	where $\lambda$ is the 4-point coupling of the Higgs field. Relevant in \eq \eqref{action_equivalent_metric} is the potential after the conformal transformation:
	\begin{equation} \label{inflationaryPotential}
		U(h) \equiv \frac{V(h)}{\Omega^4} \approx \frac{\lambda M_P^4}{4 \xi^2} \left(1 - \frac{2 M_P^2}{\xi h^2}\right) \;,
	\end{equation}
	where as before (see \eq \eqref{selec_rule_3})
	\begin{equation}
		\Omega^2 = 1 +\frac{\xi h^2 }{M_P^2}\;,
	\end{equation} 
	and we restored factors of $M_P$.
	In the second equality of \eq \eqref{inflationaryPotential}, we assumed
	\begin{equation} \label{approximationInflation}
		\xi \gg 1 \;, \qquad h \gtrsim M_P/\sqrt{\xi} \;,
	\end{equation}
	and we shall stick to this approximation in the following. 
	We observe that $U(h)$ develops a plateau for large values of $h$, which is suitable for inflation.
	
	\subsection{Review of previous results}
	To begin with, we will briefly review known results.\footnote
	{Detailed reviews of metric and Palatini Higgs inflation can be found in \cite{Rubio:2018ogq} and \cite{Tenkanen:2020dge}, respectively. We remark that we shall restrict ourselves to a classical analysis in the following. It is known, however, that in certain cases quantum effects can significantly alter the predictions of Higgs inflation; see in particular \cite{Barvinsky:2008ia,Bezrukov:2008ej,DeSimone:2008ei,Burgess:2009ea,Barbon:2009ya,Bezrukov:2009db, Barvinsky:2009fy, Barvinsky:2009ii, Bezrukov:2010jz,Hamada:2014iga,Bezrukov:2014bra,Bezrukov:2014ipa, Fumagalli:2016lls, Enckell:2016xse,Escriva:2016cwl,Bezrukov:2017dyv} for studies in the metric case and \cite{Bauer:2010jg,Rasanen:2017ivk, Markkanen:2017tun,Enckell:2018kkc,Jinno:2019und,Shaposhnikov:2020fdv,Enckell:2020lvn} for investigations including the Palatini scenario.}
	Originally \cite{Bezrukov:2007ep}, Higgs inflation was proposed in the metric version of GR. Since both torsion and non-metricity are assumed to vanish in this formulation, this corresponds to setting $A_i = B_i = C_i = D_i = E_i = 0$ and $\tilde{K} =1$ in our result \eqref{230}. We obtain as coefficient for the kinetic term of the Higgs field
	\begin{equation} \label{Kmetric}
		K(h)_{\text{metric}}=\frac{1}{\Omega^2}\left[1 + \frac{6\xi^2h^2}{M_P^2\Omega^2}\right] \approx \frac{6 M_P^2}{h^2} \;,
	\end{equation}
	where as before we used in the second step that $h$ is large. We note that $K(h)$ is non-trivial solely because of the conformal transformation \eqref{158}. The potential \eqref{inflationaryPotential} together with the kinetic term \eqref{Kmetric} define the model of metric Higgs inflation. One can test it by deriving observables in the cosmic microwave background (CMB). Important are the spectral index $n_s$, where $n_s -1$ describes the breaking of scale invariance in the spectrum of scalar perturbations, and the tensor-to-scalar ratio $r$, which determines the amplitude of primordial gravitational waves relative to scalar perturbations. The parameter $n_s$  has been measured precisely \cite{Planck:2018jri} while we only have an upper bound on $r$ \cite{Planck:2018jri, BICEP:2021xfz}. 
	We shall not repeat the analysis of metric Higgs inflation but simply quote the results of \cite{Bezrukov:2007ep}:
	\begin{equation} \label{observablesMetric}
		\left(n_s\right)_{\text{metric}} = 1 - \frac{2}{N} \;, \qquad 	\left(r\right)_{\text{metric}} = \frac{12}{N^2} \;,
	\end{equation}
	where $50 \lesssim N \lesssim  60$ sets the number of $e$-foldings before the end of inflation at which CMB observables are generated. Moreover, the observed amplitude of fluctuations determines that the non-minimal coupling $\xi$ lies between $\sim 5 \cdot 10^2$ and $\sim 5 \cdot 10^3$. This uncertainty in $\xi$ is due to the fact that we do not know the value of the quartic coupling $\lambda$ at high energies (see \eg \cite{Shaposhnikov:2020fdv}). For a given $\lambda$, however, all parameters in the model are uniquely determined. 
	Since $\xi$ is large, the approximation \eqref{approximationInflation} is well justified \cite{Bezrukov:2007ep}. The predictions \eqref{observablesMetric} agree excellently with current observations \cite{Planck:2018jri, BICEP:2021xfz}.
	
	Soon after the original proposal \cite{Bezrukov:2007ep}, a second version of Higgs inflation was developed \cite{Bauer:2008zj} in the Palatini formulation of gravity. In the terminology of the present paper, this corresponds to a special case of Weyl gravity in which the purely gravitational part of the action only consists of the Ricci scalar $R$. Correspondingly, the conformal factor $\Omega^2$  only couples to $R$ and we obtain the action:
	\begin{equation}
		\begin{aligned}
			S=&\int \mathrm{d}^{4} x \sqrt{-g}[\frac{1}{2} \Omega^2(h) R-\frac{1}{2} g^{\alpha \beta} \partial_\alpha h \partial_\beta h-V(h)] \;.
		\end{aligned}
		\label{action_Palatini}
	\end{equation}
	Using \eq \eqref{curvatureSplit}, we can decompose $R$ into its Levi-Civita part $\mathring{R}$ and contributions due to non-metricity. In the action \eqref{general_action_component}, this leads to 
	\begin{subequations} \label{substitutionsPalatini}
		\begin{align}
			&A_3 = - \frac{1}{2}\Omega^2 \;, \quad A_4 = \frac{1}{2}\Omega^2 \;,\quad B_1 = - \frac{11}{144}\Omega^2 \;,\quad B_2 =  \frac{1}{36}\Omega^2 \;,\quad B_3 =  \frac{1}{9}\Omega^2 \;,\\
			& B_4 =  \frac{1}{8}\Omega^2 \;,\quad B_5 = - \frac{1}{4}\Omega^2\;,\quad	A_1=A_2=C_i=D_i=E_i=0 \;, \quad \tilde{K}=1 \;,
		\end{align}
	\end{subequations}
	where we also imposed the vanishing of torsion. Plugging this in our result shown in \eq \eqref{230}, we obtain the kinetic term
	\begin{equation} \label{KPalatini}
		K(h)_{\text{Palatini}}=\frac{1}{\Omega^2} \approx \frac{M_P^2}{\xi h^2} \;,
	\end{equation}
	where again we used in the second step that $h$ is large. At first sight, it may seem surprising that the intricate substitutions \eqref{substitutionsPalatini} lead to the simple kinetic term \eqref{KPalatini}. As is well-known \cite{Bauer:2008zj}, however, there is a simpler way to derive the result \eqref{KPalatini}. Namely, one can immediately perform the conformal transformation $g_{\alpha \beta} \rightarrow \Omega^{-2} g_{\alpha \beta}$ in \eq \eqref{action_Palatini}. Since $R_{\mu\nu}$ is independent of $g_{\mu\nu}$ in a first-order formalism, the rescaling of the metric is easy to perform. Afterwards it becomes evident that non-metricity vanishes and we obtain the kinetic term \eqref{KPalatini}. Together with the potential \eqref{inflationaryPotential}, it defines the model of Palatini Higgs inflation. Again we quote the results for the spectral index and the tensor-to-scalar ratio \cite{Bauer:2008zj}:
	\begin{equation} \label{observablesPalatini}
		\left(n_s\right)_{\text{Palatini}} = 1 - \frac{2}{N} \;, \qquad 	\left(r\right)_{\text{Palatini}} = \frac{2}{\xi N^2} \;,
	\end{equation}
	where $\xi$ lies in the range between $\sim 10^6$ and $\sim 10^8$ in the Palatini case. Comparing with \eq \eqref{observablesMetric}, we observe that the formula for the spectral index is identical to the metric scenario but the tensor-to-scalar ratio is significantly smaller. We remark, however, that the numerical values of the spectral index do not coincide in the two models. The reason is that $N$ depends on  how inflation ends, \ie the properties of preheating and in particular the preheating temperature. Since they are different in the two cases \cite{Ema:2016dny,DeCross:2016cbs, Rubio:2019ypq, Cheong:2021kyc, Dux:2022kuk}, $N$ is slightly smaller in the Palatini scenarios (see \cite{Rubio:2019ypq} for a detailed comparison). Also the predictions of Palatini Higgs inflation are in excellent agreement with current observations of the CMB \cite{Planck:2018jri, BICEP:2021xfz}.

	In the metric and Palatini scenarios of Higgs inflation, uniqueness of predictions -- as shown in \eqs \eqref{observablesMetric} and \eqref{observablesPalatini} -- is achieved since there is only a single free parameter $\xi$ in the model. It is fixed by the requirement of matching the amplitude of scalar perturbations observed in the CMB. In other formulations of GR, however, more than one a priori unknown coupling constant emerges when coupling the Higgs field non-minimally to gravity. Correspondingly, Higgs inflation no longer leads to unique predictions beyond the special cases of the metric and Palatini scenarios \cite{Rasanen:2018ihz, Raatikainen:2019qey, Langvik:2020nrs, Shaposhnikov:2020gts}. In particular, it is also possible that the spectral index deviates from $1-2/N$ for some choices of coupling constants \cite{Rasanen:2018ihz, Raatikainen:2019qey, Langvik:2020nrs, Shaposhnikov:2020gts}.
	
	\subsection{Findings in generic metric-affine formulation}
	Evidently, the space of possible scenarios increases even more in the generic metric-affine model that we consider, which features $9$ independent parameters (see \eq \eqref{kinetic_after_conformal}). While we leave a more comprehensive study of inflationary dynamics in this model for future work, we shall briefly point out that certain regions of parameter space still reproduce the predictions of metric and Palatini Higgs inflation.
	
	First, we consider the case in which the non-minimal coupling of the Higgs field and gravity only happens through the full Ricci scalar $R$. Thus, we consider the action \eqref{action_Palatini} but now in the metric-affine formulation, in which both torsion and non-metricity are present. In \eq \eqref{general_action_component}, this corresponds to 
	\begin{subequations} \label{substitutionsPalatiniMetricAffine}
		\begin{align}
			&A_1 = 0 \;, \quad A_2 = \Omega^2 \;, \quad A_3 = - \frac{1}{2}\Omega^2 \;, \quad A_4 = \frac{1}{2}\Omega^2 \;,\\
			& B_1 = - \frac{11}{144}\Omega^2 \;,\quad B_2 =  \frac{1}{36}\Omega^2 \;,\quad B_3 =  \frac{1}{9}\Omega^2 \;, \quad B_4 =  \frac{1}{8}\Omega^2 \;,\quad B_5 = - \frac{1}{4}\Omega^2\;,\\
			& C_1=-\frac{1}{3}\Omega^2\;, \quad C_2=\frac{1}{48}\Omega^2\;, \qquad C_3=0 \;, \qquad C_4=\frac{1}{4}\Omega^2\;,\\
			& E_1=-\frac{1}{3} \Omega^2 \;, \quad E_3=\frac{1}{3}\Omega^2 \;,\quad E_5=\frac{1}{2} \Omega^2 \;, \quad D_i = E_2 = E_4 =0\;, \quad \tilde{K}=1 \;.
		\end{align}
	\end{subequations}
	When we plug this in our result \eq \eqref{230}, we again obtain the kinetic function \eqref{KPalatini}:
	\begin{equation} 
		K(h)=\frac{1}{\Omega^2} \;.
	\end{equation}
	Consequently, only allowing for a non-minimal coupling to the full Ricci scalar $R$ still leads to the predictions of Palatini Higgs inflation, which are shown in \eq \eqref{observablesPalatini}. Evidently, one arrives at almost identical observables even if the other coupling constants do not vanish exactly. As long as they are sufficiently small and the effect of $\Omega^2$ still dominates, \eqs \eqref{KPalatini} and \eqref{observablesPalatini} remain good approximations.

	Alternatively, one could consider a generic situation in which all parameters in the kinetic function $K(h)$ are large. Since in general large values of $h$ are relevant for inflation, we can try to consider the limit $h\rightarrow \infty$. Such an approach, which was already suggested in  \cite{Rasanen:2018ihz,Karananas:2020qkp}, yields in \eq \eqref{kinetic_after_conformal}
	\begin{equation} \label{KAsymptotic}
		K(h)\approx\frac{M_P^2}{\xi h^2}\left[\frac{P_7}{O_4^2} + 6\xi \right] \;.
	\end{equation}
	If we now consider the case $|P_7|/O_4^2 \lesssim 6\xi$ , we reproduce the kinetic term \eqref{Kmetric} of the metric case, which leads to the predictions \eqref{observablesMetric}. Thus, one can reproduce the observables of metric Higgs inflation even if many large observables are present in our generic metric-affine theory. We must remark, however, that in general an argument based on the limit $h\rightarrow \infty$ is not suited for deriving inflationary predictions. The reason is that CMB observables are generated at a finite value of $h$, which corresponds to $N$ $e$-foldings before the end of inflation. For example, it was shown explicitly in \cite{Shaposhnikov:2020gts} that the form of $K(h)$ at the time when CMB perturbations are generated can differ significantly from its asymptotic form achieved for $h\rightarrow \infty$. 
	
	Finally, it is interesting to note that the asymptotic behavior of the kinetic function $K(h)$ as shown in \eq \eqref{KAsymptotic} also exists in the Einstein-Cartan formulation (or equivalently Weyl gravity). For $h\rightarrow \infty$, the corresponding kinetic function \eqref{KEC} yields: 
	\begin{equation}
		K(h)_\text{EC}\approx\frac{M_P^2}{\xi h^2}\left[\frac{8\tilde{H}_1}{\tilde{O}_2} + 6\xi \right] \;,
	\end{equation}
	which coincides with \eq \eqref{KAsymptotic} upon identifying $P_7/O_4^2 \leftrightarrow  8\tilde{H}_1/\tilde{O}_2$. 
	
	\section{Conclusion}
	\label{sec:conclusion}
	An inherent ambiguity exists in GR because of its different formulations. They are all equivalent in pure gravity but can lead to distinct observable predictions once GR is coupled to matter. Since so far there is no compelling experimental or observational evidence that would favor any of the options, one can look for conceptual arguments to single out a particular version of GR. For example, it is interesting to ask which formulation can be regarded as the simplest one. Two possible answers are the following.
	
	First, one can try to minimize the number of fundamental fields or select the least involved geometry. Arguably, both conditions are fulfilled by the most commonly-used metric formulation, in which the metric is the only independent degree of freedom and the connection is uniquely determined by the requirement that torsion and non-metricity vanish. However, there are different possibilities to fulfill these conditions. In the purely-affine version of GR, one fundamental field is sufficient, namely the connection. Moreover, teleparallel formulations, in which curvature is assumed to vanish, lead to simple geometries too. 
	
	A second option exists in the quest for simplicity: One can try to minimize the number of assumptions. This singles out the metric-affine formulation, in which one does not require a priori that curvature, torsion or non-metricity are absent. Instead, all these three geometric properties are determined dynamically through the principle of stationary action. In pure gravity, this leads to vanishing torsion and non-metricity so that metric-affine gravity becomes indistinguishable from the metric formulation of GR. Once matter is included, however, torsion and non-metricity can be sourced and this equivalence is generically broken. In all cases, metric-affine gravity does not feature additional propagating degrees of freedom beyond the two polarizations of the massless graviton.
	
	The goal of the present paper was to advance the study of metric-affine gravity. Specializing to the example of a scalar field, we first constructed a general action for coupling GR to matter. In doing so, our guideline was to include all terms that are on the same footing as the non-minimal coupling to curvature, which already exists in metric GR. This led to $39$ a priori undetermined coupling constants. Subsequently, we solved for torsion and non-metricity. Plugging the results back into the original action, we derived an equivalent theory in the metric formulation of GR, in which effects of torsion and non-metricity are replaced by a specific set of higher-dimensional operators in the matter sector. For a scalar field, they can be mapped to modifications of the kinetic term. Our model encompasses the metric, Palatini, Einstein-Cartan and Weyl formulations as special cases. Moreover, we pointed out a new symmetry between the Einstein-Cartan and Weyl versions of GR.
	
	The presence of additional coupling constants is not necessarily a desirable feature because it leads to a loss of predictivity.
	However, it is forced upon us by the fact that GR exists in different formulations. Even if we want to stay as close as possible to metric GR, we have to consider at the very least all theories that are equivalent in pure gravity. The presence of undetermined parameters is a direct consequence of this inherent ambiguity of GR. Of course, it is possible to assume that these coupling constants vanish, \eg by imposing that torsion and non-metricity are absent.\footnote{Another possibility to fix some of the free coefficients is to impose a local scale symmetry \cite{Karananas:2021gco}.} Since the different non-minimal coupling parameters appear to be on the same footing, the most obvious choice would be to demand that all of them, including the non-minimal coupling to the Ricci scalar, vanish or are sufficiently small.

	While such an assumption is certainly worth exploring, it would lead to severe constraints. As a famous example, it would be incompatible with the proposal of Higgs inflation \cite{Bezrukov:2007ep}, which is only phenomenologically viable if a large non-minimal coupling to gravity exists. In the original model \cite{Bezrukov:2007ep}, which employed the metric formulation, a coupling to the Ricci scalar was considered but many more possibilities exist beyond the special case of metric GR \cite{Bauer:2008zj,Rasanen:2018ihz, Raatikainen:2019qey, Langvik:2020nrs, Shaposhnikov:2020gts}. Generically, this spoils the uniqueness of predictions and makes it necessary to systematically investigate how observables depend on the formulation of gravity. The present paper lays the groundwork for such a study of Higgs inflation in metric-affine gravity. As an outlook, we have pointed out that the predictions of the metric \cite{Bezrukov:2007ep} and Palatini \cite{Bauer:2008zj} scenarios are recovered in certain regions of parameter space.
	
	So far, only the very first steps have been taken in exploring the phenomenological consequences of metric-affine GR. Firstly, a more complete study of Higgs inflation remains to be performed. Secondly, it would be very interesting to go beyond the special case of a scalar field and include other forms of matter. As a particular example, this can have important consequences for fermions as dark matter candidates \cite{Shaposhnikov:2020aen}. We hope to report on some of these points in the future.
	
	\acknowledgments
	
	It is a pleasure to thank Syksy Räsänen and Misha Shaposhnikov for discussions and important feedback on the paper as well as Georgios Karananas for useful comments on the manuscript. This work was supported by ERC-AdG-2015~grant~694896.
	
	\appendix
	\section{Useful formulas}  \label{app:formulas}
	In this appendix, we present a few useful formulas. 
	
	If we split a general Christoffel symbol as $\Gamma^\gamma_{~\alpha \beta}= \mathring{\Gamma}^\gamma_{~\alpha \beta}+C^{\gamma}_{~\alpha \beta}$, where $\mathring{\Gamma}^\gamma_{~\alpha \beta}$ corresponds to the Levi-Civita connection, then we can decompose the Riemann tensor (see \eq \eqref{curvature_general}) as follows: 
	\begin{equation}
		R^\alpha_{~\mu\rho\nu} = \mathring{R}^\alpha_{~\mu\rho\nu} + \mathring{\nabla}_\rho C^\alpha_{~\nu\mu} -  \mathring{\nabla}_\nu C^\alpha_{~\rho\mu} + C^\alpha_{~\rho\lambda} C^\lambda_{~\nu\mu} - C^\alpha_{~\nu\lambda}C^\lambda_{~\rho\mu} \;.
	\end{equation}
	Here $\mathring{R}^\alpha_{~\mu\rho\nu}$ is the Riemann tensor defined in terms of $\mathring{\Gamma}^\gamma_{~\alpha \beta}$ (see \eq \eqref{riemannLeviCivita}) and analogously $\mathring{\nabla}_\rho$ is the covariant derivative of the Levi-Civita connection. Expanding $R^\alpha_{~\mu\rho\nu}$ explicitly in terms of contorsion and disformation gives: 
	\begin{equation}
		\begin{split}
			R^{\lambda}_{~\tau \alpha \beta}& =\mathring{R}^{\lambda}_{~\tau \alpha \beta}+2\mathring{\nabla}_{[\alpha}K^{\lambda}_{~\beta] \tau }+2\mathring{\nabla}_{[\alpha}J^{\lambda}_{~\beta] \tau }\\
			&+2 K^\lambda_{~[\alpha| \gamma }K^\gamma_{~\beta] \tau }+
			2 J^\lambda_{~[\alpha| \gamma }J^\gamma_{~\beta] \tau }
			+2 K^{\lambda}_{~[\alpha|\gamma }J^\gamma_{~\beta]\tau }+2 J^\lambda_{~[\alpha| \gamma }K^\gamma_{~\beta] \tau } \;.
			\label{split_riemann}
		\end{split}
	\end{equation}
	
	We evaluate quadratic terms composed of the full torsion- and non-metricity tensors in terms of the vector- and tensor-contributions defined in \eqs \eqref{torsionTrace} to \eqref{irrep_metricity}:
	\begin{align}
		Q_{\alpha \beta \gamma}Q^{\alpha \beta \gamma} &= \frac{1}{18}(5Q^\alpha Q_\alpha + 8\hat{Q}^\alpha \hat{Q}_\alpha-4\hat{Q}_\alpha Q^\alpha) + q_{\alpha \beta \gamma}q^{\alpha \beta \gamma}\;, \label{Qdecomposition1} \\
		Q_{\alpha \beta \gamma}Q^{\gamma \alpha \beta} &=  \frac{1}{18}(- Q^\alpha Q_\alpha + 2\hat{Q}^\alpha \hat{Q}_\alpha+8\hat{Q}_\alpha Q^\alpha )+ q_{\alpha \beta \gamma}q^{\gamma \alpha \beta}\;, \label{Qdecomposition2} \\
		T_{\alpha \beta \gamma}T^{\alpha \beta \gamma} &=\frac{2}{3}T_\alpha T^\alpha-\frac{1}{6}\hat{T}_\alpha \hat{T}^\alpha +t_{\alpha \beta \gamma}t^{\alpha \beta \gamma} \;,\\
		T_{\beta \alpha \gamma}T^{\gamma \beta \alpha} &=-\frac{1}{3}T_{\alpha}T^\alpha-\frac{1}{6}\hat{T}_\alpha \hat{T}^\alpha+t_{\beta \alpha \gamma} t^{\gamma \alpha \beta} \;, \\
		Q^{\gamma \alpha \beta}T_{\alpha \gamma \beta} &=\frac{1}{3}T_\alpha(Q^\alpha -\hat{Q}^\alpha)+q^{\gamma \alpha \beta}t_{\alpha \gamma \beta} \;. \label{QTdecomposition}
	\end{align}

	\section{Parallel transport along closed curved: effects of torsion and non-metricity}
	\label{app:parallel_transport}
	In this appendix, we shall explicitly demonstrate how torsion and/or non-metricity affect the parallel transport of a vector $v^\alpha$ along an infinitesimal closed path. Using $\tau$ as affine parameter parametrizing the path, we get for the change $\Delta v^\alpha$ of the vector (see \eg \cite{Weinberg:1972kfs}): 
	\begin{equation} \label{changeVector}
		\begin{split}
			\Delta v^\alpha &=\oint \diff \tau \frac{\diff v^\alpha}{\diff \tau} \\ &= -\oint \diff \tau \Gamma^\alpha_{~\beta \gamma} \frac{\diff x^\beta}{\diff \tau} v^\gamma \;,
		\end{split}
	\end{equation}
	where we used \eq \eqref{parallel_transport}.
	We can Taylor expand the vector and the connection around the origin: 
	\begin{equation}
		\Gamma^\alpha_{~\beta \gamma}(x)= \Gamma^\alpha_{~\beta \gamma}(0)+\partial_\nu \Gamma^\alpha_{~\beta \gamma}\rvert_{_0} x^\nu + \mathcal{O}(x^2) \;,
	\end{equation}
	\begin{equation}
		\begin{split}
			v^\gamma(x)&=v^\gamma(0)+\frac{\diff v^\gamma}{\diff x^\nu} \rvert_{_0} x^\nu + \mathcal{O}(x^2) \\ &=v^\gamma(0)-\Gamma^\gamma_{~\nu \rho} v^\rho\rvert_{_0} x^\nu + \mathcal{O}(x^2) \;.
		\end{split}
	\end{equation}
	Plugging this in \eq \eqref{changeVector} and dropping $\mathcal{O}(x^3)$ terms, we obtain: 
	\begin{equation}
		\Delta v^\alpha=  \oint \diff \tau \frac{\diff x^\beta}{\diff \tau} (\Gamma^\alpha_{~\beta \gamma}(0)\Gamma^{\gamma}_{~\nu \rho}(0)v^\rho(0)-v^\gamma(0)\partial_\nu\Gamma^\alpha_{~\beta \gamma}\rvert_{_0} )x^\nu \;,
		\label{variation}
	\end{equation}
	where we left out the linear term because it is a total derivative. Since it follows by partial integration that $\oint \diff \tau \frac{\diff x^\beta}{\diff \tau} x^\nu = -\oint \diff \tau \frac{\diff x^\nu}{\diff\tau} x^\beta$,
	only the anti-symmetric part in $\beta$, $\gamma$ in parenthesis in \eq \eqref{variation} does not vanish, and we are left with: 
	\begin{equation}
		\Delta v^\alpha= \frac{1}{2}  \oint \diff \tau \frac{\diff x^\beta}{\diff \tau} x^\nu v^\gamma (0) 	R^\alpha_{~\gamma\beta\nu}(0) \;,
		\label{paral_transport}
	\end{equation}
	where $R^\alpha_{~\gamma\beta\nu}$  is the full Riemann tensor defined in \eq \eqref{curvature_general}. Plugging the decomposition \eqref{split_riemann} into \eq \eqref{paral_transport}, we conclude that both the Levi-Civita contribution $\mathring{R}^\alpha_{~\gamma\beta\nu}$ and torsion as well as non-metricity may induce modifications to a vector being parallel transported along a closed curve.

	\section{On independence of terms} \label{app:independence_terms}
	In \eq \eqref{general_action_component}, we did not include several non-vanishing terms involving the pure tensor part $t_{ \alpha \beta \gamma}$ of torsion. The reason is that they are proportional to other contributions already present. In the following, we shall show why this is the case. By definition (see \eq \eqref{torsionTensor}), $t_{ \alpha \beta \gamma}$ has no axial part, $\epsilon^{\alpha \beta \gamma \delta} t_{\beta \gamma \delta} =0$. This implies
	\begin{equation} \label{antisymmetrizationTorsion}
		t_{[\alpha \beta \gamma]} = 0 \qquad \Leftrightarrow \qquad  t_{\alpha \beta \gamma} + t_{\beta \gamma \alpha} +  t_{\gamma \alpha \beta} = 0 \;,
	\end{equation}
	where the square brackets denote full antisymmetrization. 
	
	Now we can use \eq \eqref{antisymmetrizationTorsion} to relate different terms. We begin with
	\begin{equation}
		t_{ \alpha \beta \gamma} 	t^{\alpha  \beta \gamma} = 	t_{ \alpha \beta \gamma} \left(-t^{\beta \gamma \alpha} - t^{\gamma  \alpha \beta}\right) = 2 t_{ \alpha \beta \gamma} 	t^{\beta \alpha   \gamma} \;.
	\end{equation}
	Since $t_{ \alpha \beta \gamma} 	t^{\alpha  \beta \gamma}$ is already included in \eq \eqref{general_action_component}, we do not need to consider a contribution of $t_{ \alpha \beta \gamma} 	t^{\beta \alpha   \gamma}$. 
	Next we turn to a term involving  $\epsilon_{\alpha \beta \gamma \delta}$:
	\begin{equation}
		\epsilon_{\alpha \beta \gamma \delta} \, t^{\lambda\alpha \beta } t^{\gamma \delta}_{~~\lambda} =	\epsilon_{\alpha \beta \gamma \delta} \, \left(- t^{\beta \lambda\alpha} - t^{ \alpha \beta \lambda} \right) t^{\gamma \delta}_{~~\lambda} = -2 \epsilon_{\alpha \beta \gamma \delta} \, t^{\alpha \beta \lambda} t^{\gamma \delta}_{~~\lambda}\;.
	\end{equation}
	As we already included $\epsilon_{\alpha \beta \gamma \delta} \, t^{\alpha \beta \lambda} t^{\gamma \delta}_{~~\lambda}$ in \eq \eqref{general_action_component}, we do not need to take into account $\epsilon_{\alpha \beta \gamma \delta} \, t^{\lambda\alpha \beta } t^{\gamma \delta}_{~~\lambda}$. We can reiterate this argument:
	\begin{equation}
		\epsilon_{\alpha \beta \gamma \delta}\, t_\lambda^{~\alpha \beta }t^{\lambda \gamma \delta} = 	\epsilon_{\alpha \beta \gamma \delta}\, t_\lambda^{~\alpha \beta }\left(-t^{\delta\lambda \gamma}- t^{\gamma\delta\lambda }\right) =- 2 \epsilon_{\alpha \beta \gamma \delta} \, t_\lambda^{~\alpha \beta } t^{\gamma \delta \lambda} \;,
	\end{equation}
	which shows that $\epsilon_{\alpha \beta \gamma \delta}\, t_\lambda^{~\alpha \beta }t^{\lambda \gamma \delta}$ is not independent, either.
	Finally, we have a term which also involves the pure tensor part $q_{\alpha\beta\gamma}$ of non-metricity:
	\begin{equation}
		\epsilon_{\alpha \beta \gamma \delta}\, q_{\alpha \beta}^{~~\lambda}  t_{\lambda \gamma \delta} = 	\epsilon_{\alpha \beta \gamma \delta}\, q_{\alpha \beta}^{~~\lambda} \left(-t_{\delta\lambda \gamma} -t_{\gamma\delta\lambda }\right) = -2 	\epsilon_{\alpha \beta \gamma \delta}\, q_{\alpha \beta}^{~~\lambda} t_{\gamma\delta\lambda } \;.
	\end{equation}
	As $\epsilon_{\alpha \beta \gamma \delta}\, q_{\alpha \beta}^{~~\lambda} t_{\gamma\delta\lambda }$ is already included in \eq \eqref{general_action_component}, we can omit $\epsilon_{\alpha \beta \gamma \delta}\, q_{\alpha \beta}^{~~\lambda}  t_{\lambda \gamma \delta}$.
	Finally, one can wonder if an analogous argument can be applied to terms involving only the non-metricity tensor $q_{\alpha\beta\gamma}$. The answer is negative and the reason is that $q_{\alpha\beta\gamma}$ does not fulfill any (anti-)symmetry property (see \eq \eqref{nonMetricityTensor}).

	\bibliographystyle{JHEP}
	\bibliography{EC}
	
\end{document}